\documentclass[12pt, draftclsnofoot, onecolumn]{IEEEtran}
\usepackage{cite}
\usepackage{amsfonts}
\usepackage{times}
\usepackage{graphicx}
\usepackage{latexsym}
\usepackage{dsfont}
\usepackage{amssymb}
\usepackage{amsmath}
\usepackage{cite}
\usepackage{verbatim}
\usepackage{subfigure}

\def\bb0{{\mathbb{0}}}

\def\ba{{\mathbf{a}}}
\def\bb{{\mathbf{b}}}

\def\bff{{\mathbf{f}}}

\def\bh{{\mathbf{h}}}

\def\b0{{\mathbf{0}}}

\def\bbE{{\mathbb{E}}}

\def\cF{\mathcal{F}}

\def\sf0{{\mathsf{0}}}

\def\vec{\mathrm{vec}~}

\usepackage{graphicx}
\usepackage{array}
\usepackage{tikz}
\usepackage{tikz-3dplot}
\usetikzlibrary{shapes.geometric}
\usetikzlibrary{decorations.pathreplacing}
\usetikzlibrary{positioning}
\usepackage{pgfplots}
\usepackage{pstool}  
\usepackage{tikz}
\usetikzlibrary{automata,positioning}

\usepackage[linesnumbered,algoruled,boxed,lined]{algorithm2e}
\usepackage[absolute]{textpos}
\usepackage{threeparttable}
\usepackage[strict]{changepage} 
\usepackage{bm,upgreek} 
\usepackage{amssymb}
\usepackage{comment}
\usepackage{bbm}
\usepackage{hhline}
\usepackage{lipsum} 
\usepackage{enumitem}
\usepackage{soul}
\usepackage{xcolor}

\renewcommand{\thefootnote}{\fnsymbol{footnote}}

\ifCLASSINFOpdf
\else
\fi
\usepackage{amsmath}
\ifCLASSOPTIONcompsoc
  \usepackage[caption=false,font=normalsize,labelfont=sf,textfont=sf]{subfig}
\else
  \usepackage[caption=false,font=footnotesize]{subfig}
\fi
\newcommand{\given}{\,\vert\,}
\newcommand{\Given}{\,\bigg\vert\,}
\newcommand{\review}[1]{{\color{black}{#1}}}
\newcommand{\reviewsecond}[1]{{\color{black}{#1}}}
\newcommand{\reviewthird}[1]{{\color{black}{#1}}}

\pgfplotsset{compat=1.14}

\begin{document}
%
\title{Deep Reinforcement Learning for 5G Networks: Joint Beamforming, Power Control, and Interference Coordination}
%

\author{Faris~B.~Mismar,~\IEEEmembership{Senior~Member,~IEEE,}%
        ~Brian~L.~Evans,~\IEEEmembership{Fellow,~IEEE,}%
        ~and \\ Ahmed Alkhateeb,~\IEEEmembership{Member,~IEEE}
        
\thanks{F. B. Mismar and B. L. Evans are with the Wireless Networking and Communications Group, Dept. of Electrical and Comp. Eng., The University of Texas at Austin, Austin, TX, 78712, USA. e-mail: faris.mismar@utexas.edu, bevans@ece.utexas.edu. A. Alkhtaeeb is with the School of Electrical, Computer and Energy Engineering at Arizona State University, Tempe, AZ 85287, USA. email: alkhateeb@asu.edu.} %
\thanks{A preliminary version of this work was presented at the 2018 Asilomar Conf. on Signals, Systems and Computers \cite{8645168}.}
}

\maketitle

\begin{abstract}
The fifth generation of wireless communications (5G) promises massive increases in traffic volume and data rates, as well as improved reliability in voice calls.  Jointly optimizing beamforming, power control, and interference coordination in a 5G wireless network to enhance the communication performance to end users poses a significant challenge.  In this paper, we formulate the joint design of beamforming, power control, and interference coordination as a non-convex optimization problem to maximize the signal to interference plus noise ratio (SINR) and solve this problem using deep reinforcement learning.  By using the greedy nature of deep Q-learning to estimate future rewards of actions and using the reported coordinates of the users served by the network, we propose an algorithm for voice bearers and data bearers in sub-6 GHz and millimeter wave (mmWave) frequency bands, respectively.  The algorithm improves the performance measured by SINR and sum-rate capacity.  In realistic cellular environments, the simulation results show that our algorithm outperforms the link adaptation industry standards for sub-6 GHz voice bearers. \reviewthird{For data bearers in the mmWave frequency band, our algorithm approaches the maximum sum rate capacity, but with less than 4\% of the required run time.}
\end{abstract}

%
\IEEEpeerreviewmaketitle

\section{Introduction}\label{sec:intro} 
%
%
%
%



 

The massive growth in traffic volume and data rate continues to evolve with the introduction of {fifth generation of wireless communications} (5G).  Also evolving is enhanced voice call quality with better reliability and improved codecs.   Future wireless networks are therefore expected to meet this massive demand for both the data rates and the enhanced voice quality.  In an attempt to learn the implied characteristics of inter-cellular interference and inter-beam interference, we propose an online learning based algorithm based on a {reinforcement learning} (RL) framework.  We use this framework to derive a near-optimal policy to maximize the end-user {signal to interference plus noise} (SINR) and sum-rate capacity.  The importance of reinforcement learning in power control has been demonstrated in \cite{5683371, 8645168, 5700414}.  Power control in voice bearers makes them more robust against wireless impairments, such as fading.  It also enhances the usability of the network and increases the cellular capacity.    For data bearers, beamforming, power control, and interference coordination, can  improve the robustness of these data bearers, improve the data rates received by the end-users, and avoid retransmissions.

\review{A major question here is whether there exists a method that (1) can jointly solve for the power control, interference coordination, and beamforming, (2) achieve the upper bound on SINR, and (3) {avoids the exhaustive search in the action space} for both bearer types.    The aim of this paper is to propose an algorithm for this joint solution by utilizing the ability of reinforcement learning to explore the solution space by learning from interaction.  This algorithm applies to both voice and data bearers alike.  Furthermore, we study the overhead introduced as a result of passing information to a central location, which computes the solution through online learning.}

\subsection{Related Work}
Performing power control and beamforming in both uplink and downlink was studied in \cite{6782491, 8415781, 8422442, 725309}.  A jointly optimal transmit power and beamforming vector was solved for in \cite{725309} to maximize the SINR using optimization, but without regards for scattering or shadowing, which are critical phenomena in {millimeter wave} (mmWave) propagation.

The industry standards adopted the method of {almost blank subframe} (ABS) to resolve the co-channel inter-cell interference problem in LTE where two {base stations} (BSs) interfere with one another \cite{3gpp36300}.  While ABS works well in fixed beam antenna patterns, the dynamic nature of beamforming reduces the usefulness of ABS \cite{8654723}.

An online learning algorithm for link adaptation in {multiple-input multiple-output} (MIMO) bearers was studied in \cite{5683371}. The algorithm computational complexity was comparable to existing online learning approaches, but with minimal spatial overhead.  Interference avoidance in a heterogeneous network was studied in \cite{5700414}. A $Q$-learning framework for the coexistence of both macro and femto BSs was proposed.  The feasibility of decentralized self-organization of these BSs was established where the femtocells interference towards the macro BSs was mitigated.  The use of a $Q$-learning framework was also proposed in \cite{8645168}.  The framework focused on packetized voice power control in a multi-cell indoors environment.  It exploits the use of semi-persistent scheduling, which  establishes a virtual sense of a dedicated channel.  This channel enabled the power control of the downlink to ensure enhanced voice clarity compared to industry standards, which are based on fixed power allocation.


\begin{figure}[!t] 
\centering
\resizebox{0.75\textwidth}{!}{\includegraphics{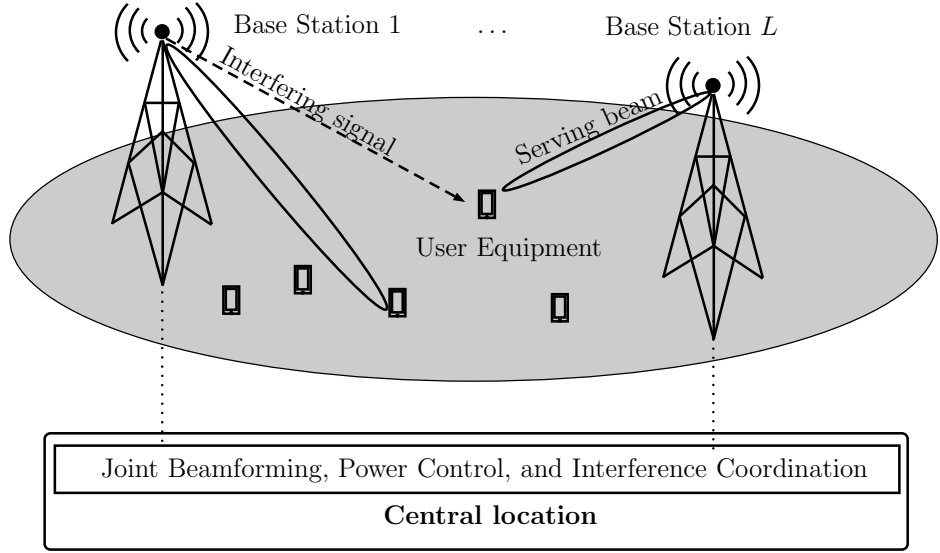}}%
\caption{Performing joint beamforming and power control on the signal from the serving base station while coordinating interference from the other BS.  \review{The decisions are computed at a central location, which can be one of the $L$ BSs.  The measurements from the UEs are relayed to the central location over the backhaul.}}
\label{fig:network}
\end{figure}


Joint power control in massive MIMO was introduced in \cite{6782491}.  This approach led to a reduced overhead due to a limited exchange of channel state information between the BSs participating in the joint power control.  The joint power control scheme led to enhanced performance measured by the SINR.  In the uplink direction, power control with beamforming was studied in \cite{8415781}.  An optimization problem was formulated to maximize the achievable sum rate of the two users while ensuring a minimal rate constraint for each user.  Using reinforcement learning to solve the problem for the uplink is computationally expensive and can cause a faster depletion of the {user equipment} (UE) battery.  We on the other hand focus on the downlink and on interference cancellation alongside power control and beamforming.

Over the last two years, the use of deep learning in wireless communications was studied in \cite{8303773, 8422442, 8645696, 8807130 ,8683468, 8335785, Alkhateeb2018, Alrabeiah19}.  The specific use of deep reinforcement learning to perform power control for mmWave was studied in \cite{8422442}. This approach was proposed as an alternative to beamforming in improving the {non-line of sight} (NLOS) transmission performance. The power allocation problem to maximize the sum-rate of UEs under the constraints of transmission power and quality targets was solved using deep reinforcement learning.  In this solution, a convolutional neural network was used to estimate the \mbox{$Q$-function} of the deep reinforcement learning problem. In \cite{8303773}, a policy that maximizes the successful transmissions in a dynamic correlated multichannel access environment was obtained using deep $Q$-learning.  The use of deep convolutional neural networks was proposed in \cite{8645696} to enhance the automatic recognition of modulation in cognitive radios at low SINRs.

\begin{table*}[!t]
\setlength\doublerulesep{0.5pt}
\caption{Literature Comparison}
\label{table:literature}
\vspace*{-1em}
\begin{adjustwidth}{-1.1cm}{0cm}
\centering
\begin{threeparttable}
\begin{tabular}{ llllll } 
\hhline{======}
Reference & Bearer & Band & Objective & Procedure\tnote{\textasteriskcentered} & Algorithm \\
\hline
\cite{6782491} & data & unspecified & downlink SINR & PC & convex optimization\\  
\cite{8415781} & data & mmWave & uplink sum-rate & BF, PC & convex optimization \\ 
\cite{8422442} & data & mmWave & dowlink SINR, sum-rate & PC & deep reinforcement learning\\ 
\cite{8807130} & data & unspecified & uplink power, sum-rate & PC & deep neural networks \\ 
\cite{8683468} & data & unspecified&  downlink throughput & PC & deep neural networks \\ 
\cite{8335785} & data & unspecified & SINR, spectral efficiency & PC & convolutional neural network \\ 
\cite{Alkhateeb2018} & data & mmWave & downlink achievable rate & BF & deep neural networks \\  
\cite{Alrabeiah19} & data & mmWave and sub-6 & downlink spectral efficiency & BF & deep neural networks \\ 
\cite{8525802} & data & unspecified & downlink sum-rate & BF & deep adversarial reinforcement learning \\ 
\cite{framework} & voice & sub-6 & downlink SINR  & PC & tabular reinforcement learning \\ 
\cite{8542687} & data & mmWave & downlink sum-rate & BF, IC & deep neural networks \\ 
\cite{Xia19} & data & unspecified & downlink SINR & BF & deep neural networks \\ 
\hline
Proposed & voice and data & mmWave and sub-6 & downlink SINR & BF, PC, IC & deep reinforcement learning \\
\hhline{======}
\end{tabular}
\begin{tablenotes}\footnotesize
\item[\textasteriskcentered]{PC is power control, IC is interference coordination, and BF is beamforming.}
\end{tablenotes}
\end{threeparttable}
\end{adjustwidth}
\end{table*}

\review{ In \cite{Alkhateeb2018}, deep neural networks were leveraged to predict mmWave beams with low training overhead using the omni-directional received signals collected from neighboring base stations. In \cite{Alrabeiah19}, the authors generalized \cite{Alkhateeb2018} by mapping the channel knowledge at a small number of antennas to an SINR-optimal beamforming vector for a larger array, even if this array was at a different frequency at a neighboring BS.}
\review{
The use of adversarial reinforcement learning in beamforming for data bearers was proposed in \cite{8525802}, where an algorithm to derive antenna diagrams with near-optimal SINR performance was devised.  There was no reference to power control or interference coordination.  Voice bearers in the sub-6 GHz frequency band was studied in \cite{framework} but only in a single co-located BS environment, in contrast with our paper where we study voice in a multi-access network with multiple BSs. Joint beamforming and interference coordination at mmWave was performed in \cite{8542687} using deep neural networks, which require knowledge of the channel to make decisions.  The performance of deep neural networks on beamforming was studied in \cite{Xia19} but without the use of reinforcement learning. Table~\ref{table:literature} shows how our work compares with earlier work.}

\subsection{Motivation}
In this paper, we provide an answer to the question whether a method exists that can perform the joint beamforming, power control, and interference coordination by introducing a different approach to power control in wireless networks.  In such a setting, it is not only the transmit power of the serving BS that is controlled as in standard implementations, but also the transmit powers of the interfering base stations {from a central location} as shown in Fig.~\ref{fig:network}.  As a result of this apparent conflict, a \textit{race condition} emerges, where the serving BS of a given user is an interfering BS of another user.   The reason why we choose deep reinforcement learning (DRL) is as follows:
\begin{enumerate}
    \item The proposed solution does not require the knowledge of the channels in order to find the SINR-optimal beamforming vector. {This is in contrast with the upper bound SINR performance, which finds the optimal beamforming vector by searching across all the beams in a codebook that maximizes the SINR (and this requires perfect knowledge of the channel)}. 
    \item The proposed solution minimizes the involvement of the UE in sending feedback to the BS.  In particular, the UE sends back its received SINR along with its coordinates, while the agent handles the power control and interference coordination commands to the involved BSs.  {Industry specifications \cite{3gpp36300} require that the UE reports its channel state information which is either a vector of length equal to the number of antenna elements or a matrix of dimension equal to the number of antenna elements in each direction.  In our case, we achieve a reduction in the reporting overhead by using the UE coordinates instead.}
    \item {The implementation complexity of upper bound SINR performance message passing for joint beamforming, power control, and interference coordination commands when multiple BSs are involved is prohibitive.}
    \item Having explicit {power control and interference coordination} (PCIC) commands sent by the UE to the serving and interfering BSs requires a modification to the current industry standards \cite{3gpp36213}.  These standards today only require the serving BS to send power control commands to the UE for the uplink direction.
\end{enumerate}

\subsection{Contributions}
In finding a different approach to power control in wireless networks, this paper makes the following specific contributions:

\begin{itemize}
    \item Formulate the joint beamforming, power control, and interference coordination problem in the downlink direction as an optimization problem that optimizes the users' received SINR.
    \item Resolve the race condition between the involved base stations in sub-exponential times in the number of antennas.  The race condition is handled by a central location (similar to coordinated multipoint \cite{8665922}) based on the user reported downlink SINR and coordinates.
    \item Show how to create a deep reinforcement learning based solution where multiple actions can be taken at once using a binary encoding of the relevant actions performed by the BS, which we define in Section~\ref{subsec:setup}.
\end{itemize}

\section{Network, System, and Channel Models}\label{sec:system}

In this section, we describe the adopted network, system, and channel models. 

\subsection{Network Model} \label{subsec:net_model}

We consider an orthogonal frequency division multiplexing (OFDM) multi-access downlink cellular network of $L$ BSs.  This network is comprised of a serving BS and at least one interfering BS.  We adopt a downlink scenario, where a BS is transmitting to one UE.  The BSs have an intersite distance of $R$ and the UEs are randomly scattered in their service area.  \reviewsecond{The association between the users and their serving BS is based on the distance between them.  A user is served by one BS maximum.}  The cell radius is $r > R/2$ to allow overlapping of coverage.  Voice bearers run on sub-6 GHz frequency bands while the data bearers use mmWave frequency band.  We employ analog beamforming for the data bearers to compensate for the high propagation loss due to the higher center frequency.

\subsection{System Model}
Considering the network model in Section \ref{subsec:net_model}, and adopting a multi-antenna setup where each BS employs a {uniform linear array} (ULA) of $M$ antennas and the UEs have single antennas, the received signal at the UE from the $\ell$-th BS can be written as 
\begin{equation} \label{eq:signal_model}
{y}_\ell = \bh^\ast_{\ell,\ell} \bff_\ell x_\ell+ \sum_{b \neq \ell} \bh^\ast_{\ell,b} \bff_b x_b + n_\ell
\end{equation}
where $x_\ell, x_b \in\mathbb{C}$ are the transmitted signals from the $\ell$-th and $b$-th BSs, and they satisfy the power constraint $\bbE [\vert x_\ell\vert^2]=P_{\mathrm{TX},\ell}$ (similarly for $b$). The $M \times 1$  vectors $\bff_\ell, \bff_b \in\mathbb{C}^{M \times 1}$ denote the adopted downlink beamforming vectors at the $\ell$-th and $b$-th BSs, while the $M \times 1$ vectors $\bh_{\ell, \ell}, \bh_{\ell, b} \in\mathbb{C}^{M \times 1}$ are the channel vectors connecting the user at the $\ell$-th BS with the $\ell$-th and $b$-th BSs, respectively. Finally, $n_\ell \sim \textrm{Normal}(0, \sigma_n^2)$ is the received noise at the user sampled from a complex Normal distribution with zero-mean and variance $\sigma_n^2$.  The first term in \eqref{eq:signal_model} represents the desired received signal, while the second term represents the interference received at the user due to the transmission from the other BSs.

\textbf{Beamforming vectors:} Given the hardware constraints on the mmWave transceivers, we assume that the BSs use analog-only beamforming vectors, where the beamforming weights of every beamforming vector $\bff_\ell, \ell=1, 2, ..., L$ are implemented using constant-modulus phase shifters, i.e., $\left[\bff_\ell\right]_m= e^{j \theta_m}$. Further, we assume that every beamforming vector is selected from a beamsteering-based beamforming codebook $\cF$ of cardinality $\left|\cF\right|:=N_{\text{CB}}$, with the $n$-th element in this codebook defined as
\begin{equation}
\begin{aligned}
    \bff_n & :=\ba(\theta_n) \\
    & = \frac{1}{\sqrt{M}} \left[1, e^{j k d \cos\left(\theta_n\right)}, ..., e^{j k d (M-1) \cos\left(\theta_n\right)}\right]^\top,
\end{aligned}
\end{equation}
where $d$ and $k$ denote the antenna spacing and the wave-number, while $\theta_n$ represents the steering angle. Finally, $\ba(\theta_n)$ is the array steering vector in the direction of $\theta_n$.  The value of $\theta_n$ is obtained by dividing the the antenna angular space between $0$ and $\pi$ radians by the number of antennas $M$.

\textbf{Power control and interference coordination:} 
Every BS $\ell$ is assumed to have a transmit power $P_{\text{TX},\ell} \in \mathcal{P}$, where $\mathcal{P}$ is the set of candidate transmit powers.   We define the set of the transmit powers as the power offset above (or below) the BS transmit power.  Our choice of the transmit power set $\mathcal{P}$ is provided in Section \ref{subsec:setup}.  This choice of $\mathcal{P}$ follows \cite{3gpp36213}.  

Power control and interference coordination take place over a semi-dedicated channel.  For voice, this is facilitated through the  semi-persistent scheduling, which creates a virtual sense of a dedicated channel as we have mentioned in Section~\ref{sec:intro}.  For data bearers, the use of beamforming provides a dedicated beam for a given UE, through which power control and interference coordination takes place.

\vspace*{-1em}
\subsection{Channel Model}
In this paper, we adopt a narrow-band geometric channel model, which is widely considered for analyzing and designing mmWave systems \cite{Alkhateeb2014,HeathJr2016,Schniter2014}. With this geometric model, the downlink channel from a BS $b$ to the user in BS $\ell$ can be written as
\begin{equation}
    \bh_{\ell,b}=\frac{\sqrt{M}}{\rho_{\ell,b}} \sum_{p=1}^{N^p_{\ell,b}}  \alpha^p_{\ell,b} \ba^\ast\left(\theta^p_{\ell,b}\right), 
    \label{eq:channel}
\end{equation}
where $\alpha^p_{\ell,b}$ and $\theta^p_{\ell,b}$ are the complex path gain and angle of departure (AoD) of the $p$-th path, and $\ba(\theta^p_{\ell,b})$ is the array response vector associated with the AoD, $\theta^p_{\ell,b}$. Note that $N^p_{\ell,b}$ which denotes the number of channel paths is normally a small number in mmWave channels compared to sub-6 GHz channels \cite{Rappaport2013,Rappaport2014}, which captures the sparsity of the channels in the angular domain. Finally, $\rho_{\ell,b}$, represents the path-loss between BS $b$ and the user served in the area of BS $\ell$. Note that the channel model in \eqref{eq:channel} accounts of both the LOS and NLOS cases. For the LOS case, we assume that $N^p_{\ell,b}=1$.

%

We define $P_\text{UE}[t]$ as the received downlink power as measured by the UE over a set of {physical resource blocks} (PRBs) at a given time $t$ as%
\begin{equation}
    P^{\ell,b}_{\text{UE}}[t]=P_{\text{TX},b}[t] \left|\bh_{\ell,b}^\ast[t] \bff_b[t]\right|^2
\end{equation}
where $P_{\text{TX},b}$ is the PRB transmit power from BS $b$. Next, we compute the received SINR for the UE served in BS $\ell$ at time step $t$ as follows:
\begin{align}\label{eq:sinr_final}
    \gamma^{\ell}[t] = \frac{P_{\text{TX},\ell}[t]  \vert \bh_{\ell,\ell}^\ast[t]  \bff_\ell[t] \vert ^2}{\sigma_n^2 + \sum_{b\neq\ell} P_{\text{TX},b} [t] \vert \bh_{\ell,b}^\ast[t] \bff_b[t]\vert ^2}.
\end{align}

\noindent This is the received SINR that we will optimize in our paper in Sections~\ref{sec:voice_pcic} and \ref{sec:bf_pcic}.
\section{Problem Formulation}\label{sec:problem}

Our objective is to jointly optimize the beamforming vectors and the transmit power at the $L$ BSs to maximize the achievable sum rate of the users.  We formulate the joint beamforming, power control, and interference coordination optimization problem as 
\begin{equation}
\label{eq:optimization}
\begin{aligned}
&  \underset{\substack{P_{\text{TX},j}[t], \ \forall j \\\bff_{j}[t], \ \forall j}}{\textrm{maximize}}  & \sum_{j\in\{1,2,\ldots,L\}} \gamma^j[t]\\
& \text{subject to} \hspace{10pt} &P_{\text{TX},j}[t] \in \mathcal{P}, \qquad \forall j, \\
&&  \bff_j[t] \in \mathcal{F},\qquad \forall j,\\
&& \gamma^{j}[t] \ge \gamma_\text{target}.
\end{aligned}
\end{equation}
where $\gamma_\text{target}$ denotes the target SINR of the downlink transmission \reviewsecond{(all SINR quantities are in dB)}.  $\mathcal{P}$ and  $\cF$ are the sets of candidate transmit powers and beamforming codebook, respectively as stated earlier. This problem is a non-convex optimization problem due to the non-convexity of the \reviewsecond{first two} constraints.  \review{The $\ell$-th BS attempts to solve this problem to find optimal $P_{\text{TX}, \ell}$ and $\mathbf{f}_\ell$ for the UE served by it at time $t$.} \review{We solve this optimization problem at a central location by searching over the space of the Cartesian product of $\mathcal{P}\times\mathcal{F}$.  The optimal solution to this problem is found through an exhaustive search over this space (i.e., by brute force).  The complexity of this search is known to be exponential in the number of BSs.  We discuss this and the overhead of the communication to a central location in Section~\ref{sec:bf_pcic}.}





Next, we provide a brief overview on deep reinforcement learning in Section~\ref{sec:deep_rl} before delving into the proposed algorithm in Sections~\ref{sec:voice_pcic} and \ref{sec:bf_pcic}.

\section{A Primer on Deep Reinforcement Learning}\label{sec:deep_rl}

\begin{figure}[!t]
\begin{adjustwidth}{0cm}{0cm}
\centering
\begin{tikzpicture}[node distance = 5em, auto, thick, scale=2, font=\scriptsize]]
    \node [rectangle, draw, 
    text width=12em, text centered, rounded corners, minimum height=2em, fill=gray!30] 
    (Agent) {JB-PCIC Algorithm (Agent)};
    \node [rectangle, draw, 
    text width=8em, text centered, rounded corners, minimum height=1em, below of=Agent] (Environment) {Cellular Network (Environment)};
    \path [draw, -latex] (Agent.0) --++ (1em,0em) |- node [text width=5em,near start,yshift=-0.1cm]{Action\\ $a\in\mathcal{A}$} (Environment.0);
    \path [draw, -latex] (Environment.180) --++ (-2em,0em) |- node [text width=5em, xshift=-0.7cm,yshift=-1.9cm] {State \\ $s\in\mathcal{S}$\\ Next State \\ $s^\prime\in\mathcal{S}$} (Agent.180);
    \path [draw, -latex] (Environment.north) --++ (0em,0em) -- node [right] {Reward $r_{s,s^\prime,a}[t;q]$} (Agent.south);
\end{tikzpicture}
\end{adjustwidth}
\vspace*{-1em}
\caption{The agent-environment interaction in reinforcement learning.}
\label{fig:rl}
\end{figure}
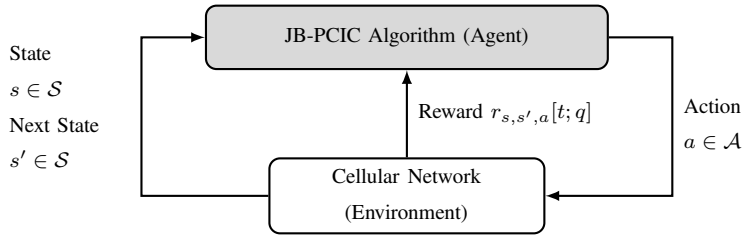

In this section, we describe {deep reinforcement learning}, which is a special type of reinforcement learning \cite{mnih2013playing}. 
Reinforcement learning is a machine learning technique that enables an \textit{agent} to discover what action it should take to maximize its expected future \textit{reward} in an interactive \textit{environment}.  The interaction between the agent and the environment is shown in Fig.~\ref{fig:rl}.  In particular, DRL exploits the ability of deep neural networks to learn better representations than handcrafted features and act as a universal approximator of functions.  

\textbf{Reinforcement learning elements:} Reinforcement learning has several elements \cite{Sutton}.  These elements interact together, and are as follows:

\begin{itemize}[leftmargin=*]
    \item \textit{Observations}: Observations are continuous measures of the properties of the environment and are written as a $p$-ary vector $\mathbf{O}\in\mathbb{R}^p$, where $p$ is the number of properties observed.
    \item \textit{States}: The state $s_t\in\mathcal{S}$ is the discretization of the observations at time step $t$.  Often, states are also used to mean observations.
    \item \textit{Actions}: An action $a_t\in\mathcal{A}$ is one of the valid choices that the agent can make at time step $t$.  The action changes the state of the environment from the current state $s$ to the target state $s^\prime$. 
    \item \textit{Policy}: A policy $\pi(\cdot)$ is a mapping between the state of the environment and the action to be taken by the agent.  We define our stochastic policy $\pi(a\given s):\mathcal{S}\times\mathcal{A}\rightarrow [0,1]$.  
    
    \item \textit{Rewards}: The reward signal $r_{s,s^\prime,a}[t;q]$ is obtained after the agent takes an action $a$ when it is in state $s$ at time step $t$ and moves to the next state $s^\prime$. \review{The parameter $q\in\{0,1\}$ is the bearer selector, which is a binary parameter to differentiate voice bearers from data bearers.}
    \item \textit{State-action value function}: The state-action value function under a given policy $\pi$ is denoted $Q_\pi(s,a)$.  It is the expected discounted reward when starting in state $s$ and selecting an action $a$ under the policy $\pi$.
\end{itemize}

These elements work together and their relationship is governed by the objective to maximize the future discounted reward for every action chosen by the agent, which causes the environment to transition to a new state.  The policy dictates the relationship between the agent and the state.  The value of the expected discounted reward is learned through the training phase.

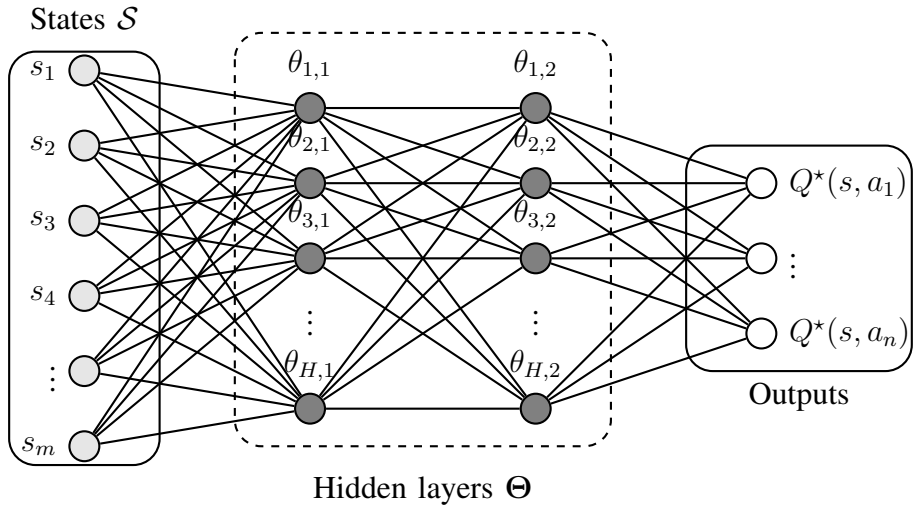
\begin{figure}[!t]
\centering
\begin{tikzpicture}[thick,scale=1, every node/.style={scale=1},   cnode/.style={draw=black,fill=#1,minimum width=3mm,circle},
]
 \node at (3,-3.75) {$\vdots$};
  \node at (6,-3.75) {$\vdots$};
    \foreach \x in {1,...,6}
    {   
   	\pgfmathparse{\x== 6 ? "s_m" : "s_\x"}
    \pgfmathparse{\x == 5 ? "\vdots" : "\pgfmathresult"}
    
        \node[cnode=gray!20,label=180:${\pgfmathresult}$] (x-\x) at (0,{-\x-int(\x/7)+0.5}) {};
    }

    \foreach \x in {1,...,4}
    {
        \pgfmathparse{\x<4 ? \x : "H"}
        \node[cnode=gray,label=90:$\theta_{\pgfmathresult,1}$] (x2-\x) at (3,{-\x-int(\x/4)}) {};
        \node[cnode=gray,label=90:$\theta_{\pgfmathresult,2}$] (p-\x) at (6,{-\x-int(\x/4)}) {};
        
    }

	\draw[rounded corners=10pt] (-1,-0.25) rectangle ++(2,-5.5) node[below] at (0, 0.5) {States $\mathcal{S}$};
	\draw[dashed, rounded corners=10pt] (2,0) rectangle ++(5,-5.5) node[below] at (4.5,-5.75) {Hidden layers $\boldsymbol{\Theta}$};
	\draw[rounded corners=10pt] (8,-1.5) rectangle ++(3,-3) node[below] at (9.5,-4.5) {Outputs};

	\foreach \x in {1,...,3}
	{
      	\pgfmathparse{\x== 3 ? "Q^\star(s, a_n)" : "Q^\star(s, a_\x)"}
      	\pgfmathparse{\x== 2? "\vdots" : "\pgfmathresult"}
     
        	\node[draw=black,circle,label=0:${\pgfmathresult}$] (s-\x) at (9,{-\x-int(\x/4)-1}) {};
	}
	    
    \foreach \x in {1,...,4}
    {   \foreach \y in {1,...,3}
        {   \draw (p-\x) -- (s-\y);
        }
    }
    
    \foreach \x in {1,...,4}
    {   \foreach \y in {1,...,4}
        {  \pgfmathparse{\x<4 ? \x : "H"}   
	        \draw (x2-\x) -- (p-\y) ; 
        }
    }
    \foreach \x in {1,...,6}
    {   \foreach \y in {1,...,4}
        {  \draw (x-\x) -- (x2-\y);
        }
    }
\end{tikzpicture}
\vspace*{-1em}
\caption{Structure of the deep $Q$-network used for the implementation of the algorithms with two hidden layers each of dimension $H$.  Here, $(u,v) = (H,2)$, $\vert\mathcal{S}\vert = m, \text{and}\, \vert\mathcal{A}\vert = n$.}
\label{fig:dqn}
\end{figure}

If $Q_\pi(s,a)$ is updated every time step, then it is expected to converge to the optimal state-action value function $Q_\pi^\star(s,a)$ as $t\to +\infty$\cite{Sutton}.  However, this may not be easily achieved. Therefore, we use a function approximator instead aligned with \cite{mnih2013playing}.  We define a neural network with its weights at time step $t$ as $\boldsymbol{\Theta}_t\in\mathbbm{R}^{u\times v}$  as in Fig.~\ref{fig:dqn}.  Also, if we define $\boldsymbol{\theta}_t := \vec\!(\boldsymbol{\Theta}_t) \in\mathbbm{R}^{uv}$, we thus build a function approximator $Q_\pi(s,a;\boldsymbol{\theta}_t) \approx Q^\star_\pi(s,a)$.  This function approximator is neural network based and is known as the {Deep Q-Network} (DQN) \cite{mnih2013playing}.  Activation functions, which are non-linear functions that compute the hidden layer values, are an important component of neural networks.  A common choice of the activation function is the sigmoid function $\sigma\colon x \mapsto 1/(1 + e^{-x})$ \cite{goodfellow}.  This DQN is trained through adjusting $\boldsymbol{\theta}$ at every time step $t$ to reduce the mean-squared error loss $L_t(\boldsymbol{\theta}_t)$:
\begin{equation}
     \underset{\boldsymbol{\theta}_t}{\textrm{minimize}}  \qquad L_t(\boldsymbol{\theta}_t) := \mathbb{E}_{s,a} \left [(y_t - Q_\pi(s,a;\boldsymbol{\theta}_t))^2 \right ] 
\end{equation}
where $y_t := \mathbb{E}_{s^\prime} [ r_{s,s^\prime,a} + \gamma \max_{a^\prime} Q_\pi(s^\prime,a^\prime;\boldsymbol{\theta}_{t-1}) \given s_t, a_t ]$ is the estimated function value at time step $t$ when the current state and action are $s$ and $a$ respectively. \review{The process of interacting with the environment and the DQN to obtain a prediction and compare it with the true answer and suffer a loss $L_t(\cdot)$ is often referred to as ``online learning.''  In online learning, the UEs feedback their data to the serving BS, which in turns relays it to the central location for DQN training.  This data represent the state of our network environment $\mathcal{S}$, as we explain further in Section~\ref{sec:jbpcic_results}.}

\reviewsecond{\textbf{DQN dimension:}  we set the dimension of the input layer in the DQN to be equal to the number of states $\vert\mathcal{S}\vert$.  The dimension of the output layer is equal to the number of actions $\vert\mathcal{A}\vert$.  For the hidden layer dimension, we choose a small depth since the depth has the greatest impact on the computational complexity.  The dimension of the width follows \cite{1189626} as we show further in Section~\ref{subsec:setup}.
}

\textbf{Deep reinforcement training phase:} In the training phase of the DQN, the weights $\boldsymbol{\theta}_t$ in the DQN are updated after every iteration in time $t$ using the {stochastic gradient descent} (SGD) algorithm on a minibatch of data.  SGD starts with a random initial value of $\boldsymbol{\theta}$ and performs an iterative process to update $\boldsymbol{\theta}$ using a step size $\eta > 0$ as follows:
\begin{align}
    \boldsymbol{\theta}_{t+1} := \boldsymbol{\theta}_t - \eta\nabla L_t(\boldsymbol{\theta}_t).
\end{align}

The training of the DQN is facilitated by ``experience replay'' \cite{lin}.  The experience replay buffer $\mathcal{D}$ stores the experiences at each time step $t$. An experience $e_t$ is defined as $e_t := (s_t, a_t, r_{s,s^\prime,a}[t;q], s^\prime_t)$.  We draw samples of experience at random from this buffer and perform minibatch training on the DQN.  This approach offers advantages of stability and avoidance of local minimum convergence \cite{mnih2013playing}.  The use of experience replay also justifies the use of off-policy learning algorithms, since the current parameters of the DQN are different from those used to generate the sample from $\mathcal{D}$.

We define the state-action value function estimated by the DQN $Q_\pi^\star(s,a)$ as
\begin{equation}
    \label{eq:bellman_deep}
    Q_\pi^\star(s_t,a_t) := \mathbb{E}_{s^\prime} \left [ r_{s,s^\prime,a} + \gamma \max_{a^\prime} Q_\pi^\star(s^\prime,a^\prime) \Given s_t, a_t \right ]\!,
\end{equation}
which is known as the Bellman equation.  Here, $\gamma\colon 0 < \gamma < 1$ is the {discount factor} and determines the importance of the predicted future rewards.  The next state is $s^\prime$ and the next action is $a^\prime$.  Our goal using DQN is to find a solution to maximize the state-action function $Q^\star_\pi(s_t,a_t)$.

Often compared with deep $Q$-learning is the tabular version of $Q$-learning \cite{Sutton}.  Despite the finite size of the states and action space, tabular $Q$-learning is slow to converge is because its convergence requires the state-action pairs to be sampled infinitely often \cite{Sutton, 5983301}.  Further, tabular RL requires a non-trivial initialization of the $\mathbf{Q}\in\mathbb{R}^{\vert \mathcal{S}\vert \times \vert \mathcal{A}\vert }$ table to avoid longer convergence times \cite{framework}.  However, deep $Q$-learning convergence is not guaranteed when using a non-linear approximator such as the DQN \cite{mnih2013playing}.  We discuss tabular $Q$-learning in Section~\ref{sec:voice_pcic}.



\textbf{Policy selection:} In general, $Q$-learning is an off-policy reinforcement learning algorithm.  An off-policy algorithm means that a near-optimal policy can be found even when actions are selected according to an arbitrary exploratory policy \cite{Sutton}.   Due to this, we choose a near-greedy action selection policy.  This policy has two modes:

\begin{enumerate}
    \item \textit{exploration}: the agent tries different actions at random at every time step $t$ to discover an effective action $a_t$.
    \item \textit{exploitation}: the agent chooses an action at time step $t$ that maximizes the state-action value function $Q_\pi(s,a;\boldsymbol{\theta}_t)$ based on the previous experience.
\end{enumerate}

In this policy, the agent performs exploration with a probability $\epsilon$ and exploitation with probability of $1-\epsilon$, where $\epsilon\colon 0 < \epsilon < 1$ is a hyperparameter that adjusts the trade-off between exploration and exploitation.  This trade-off is why this policy is also called the $\epsilon$-greedy action selection policy. \reviewthird{This policy is known to have a linear regret in $t$ (regret is the opportunity loss of one time step) \cite{silver}.}

At each time step $t$, the UEs move at speed $v$ and the agent performs a certain action $a_t$ from its current state $s_t$.  The agent receives a reward $r_{s,s^\prime,a}[t;q]$ and moves to a target state $s^\prime := s_{t+1}$. We call the period of time in which an interaction between the agent and the environment takes place an \textit{episode}. One episode has a duration of $T$ time steps.  An episode is said to have \textit{converged} if within $T$ time steps the target objective was fulfilled.

In our DQN implementation, we particularly keep track of the UE coordinates.  When UE coordinates are reported back to the network and used to make informed decisions, the performance of the network improves \cite{8403587}.  Therefore, UE coordinates need to be part of the DRL state space $\mathcal{S}$.


\section{Deep Reinforcement Learning in Voice Power Control and Interference Coordination}
\label{sec:voice_pcic}
\reviewthird{In this section, we describe our proposed voice power control and interference coordination reinforcement learning algorithm as well as the baseline solutions which we compare our solution against. First, we describe the fixed power allocation algorithm, which is the industry standard algorithm today, and then the implementation of the proposed algorithm using tabular and deep implementations of $Q$-learning.  Finally, we explain the brute force algorithm.}

\subsection{Fixed Power Allocation}
We introduce the {fixed power allocation} (FPA) power control as a baseline algorithm that sets the transmit signal power at a specific value. No interference coordination is implemented in FPA.  Total transmit power is simply divided equally among all the PRBs and is therefore constant:
\begin{equation}
P_{\text{TX}, b}[t] :=  P_\text{BS}^{\rm max} - 10\log N_\text{PRB} + 10\log N_{\text{PRB},b}[t]\qquad\text{(dBm)}  \\
\end{equation}
where $N_\text{PRB}$ is the total number of physical resource blocks in the BS and $N_{\text{PRB},b}$ is the number of available PRBs to the UE in the $b$-th BS at the time step $t$.

FPA with adaptive modulation and coding is the industry standard algorithm \cite{3gpp36213}.  \review{In this standard algorithm, the BS fixes its transmit power and only changes the} modulation and code schemes of the transmission.  This change is known as the ``link adaptation.''  Link adaptation takes place based on the reports sent by the UE back to the BS (i.e., the SINR and received power).  \review{Since the BS transmit power is fixed, the link adaptation takes place based on either periodic or aperiodic measurement feedback from the voice UE to the serving BS}.  This results in an improved effective SINR and a reduction in the voice packet error rate.   
\review{There is no measurement sent to the interfering BS based on FPA.}

\subsection{Tabular RL}
We use a tabular setting of $Q$-learning (or ``vanilla'' $Q$-learning) to implement the algorithm for voice communication.  In a tabular setting, the state-action value function $Q_\pi(s_t,a_t)$ is represented by a table $\mathbf{Q}\in\mathbb{R}^{\vert \mathcal{S}\vert\times\vert\mathcal{A}\vert}$.  There is no neural network involvement and the $Q$-learning update analog of \eqref{eq:bellman_deep} is defined as:
\begin{equation}
    Q_\pi(s_t,a_t) := (1-\alpha) Q_\pi(s_t,a_t) +  \alpha \left (r_{s,s^\prime, a} + \gamma \max_{a^\prime} Q_\pi(s^\prime, a^\prime) \right )
\end{equation}
where $Q_\pi(s_t, a_t) := [\mathbf{Q}]_{s_t,a_t}$.  Here, $\alpha > 0$ is the learning rate of the $Q$-learning update and defines how aggressive the experience update is with respect to the prior experience.  
Computationally, the tabular setting suits problems with small state spaces, and maintaining a table $\mathbf{Q}$ is possible.

\subsection{Proposed Algorithm}
We propose Algorithm~\ref{alg:algorithm} which is a DRL-based approach.  This algorithm performs both power control and interference coordination without the UE sending explicit power control or interference coordination commands.  This use of the DQN may provide a lower computational overhead compared to the tabular $Q$-learning depending on the number of states and the depth of the DQN \cite{framework}.   The main steps of Algorithm~\ref{alg:algorithm} are as follows:
\begin{itemize}
    \item Select an optimization action at a time step $t$.
    \item Select a joint beamforming, power control, and interference coordination action.
    \item Assess the impact on \reviewsecond{effective SINR} $\gamma_\text{eff}^\ell[t]$.
    \item Reward the action taken based on the impact on $\gamma_\text{eff}^\ell[t]$ and its distance from $\gamma_\text{target}$ or $\gamma_\text{min}$.
    \item Train the DQN based on the outcomes.
\end{itemize}

Power control for the serving BS $b$ is described as
\begin{equation}
    P_{\text{TX}, b}[t] = \min (P_\text{BS}^{\rm max},P_{\text{TX}, b}[t - 1] + \text{PC}_b[t]).
    \label{eq:pc}
\end{equation}

\noindent We add one more condition for the interference coordination on the interfering BS $\ell$ as
\begin{equation}
    \begin{aligned}
        &P_{\text{TX}, \ell}[t] = \min (P_\text{BS}^{\rm max},P_{\text{TX}, \ell}[t - 1] + \text{IC}_\ell[t]) \\
    \end{aligned}
    \label{eq:ic}
\end{equation}
where the role of the BS (serving vs.\ interfering) can change based on the UE being served.  IC and PC commands are actually the same, but the role of the BS makes one an interferer (which needs coordination) and the other a server (which needs power control).
We model the PCIC algorithm using deep $Q$-learning as shown in Algorithm~\ref{alg:algorithm}.   Our proposed algorithm solves (\ref{eq:optimization}).

Different from \cite{8645168}, we use the effective SINR $\gamma_\text{eff}^\ell[t]$ \reviewsecond{(i.e., the SINR including coding gain)} for all three voice algorithms where the adaptive code rate $\beta$ is chosen based on the SINR $\gamma^\ell[t]$.  We use an {adaptive multirate} (AMR) codec and quadrature phase shift keying modulation for voice. %
We choose to fix the modulation since voice bearers do not typically require high data rates \cite{8645168}.  This effective SINR $\gamma_\text{eff}^\ell[t]$ is the quantity we optimize in Algorithm~\ref{alg:algorithm}.
 
For FPA, the run-time complexity is $\mathcal{O}(1)$.  For tabular \mbox{$Q$-learning} PCIC, the run-time complexity is $\mathcal{O}(\vert\mathcal{S}^\textrm{voice}\vert \vert\mathcal{A}^\textrm{voice}\vert)$\cite{framework}, where $ \mathcal{S}^\textrm{voice},\mathcal{A}^\textrm{voice}$ are the state and action sets for voice bearers.

\review{Since one of the $L$ BSs also serves as a central location to the surrounding BSs in our proposed algorithm, the overhead due to transmission over the backhaul to this central location for a total of $N_\text{UE}$ UEs in the service area is in $\mathcal{O}(gLN_\text{UE})$, where the periodicity $g$ is the number of measurements sent by any given UE during time step $t$ \cite{3gpp38211}.}
 

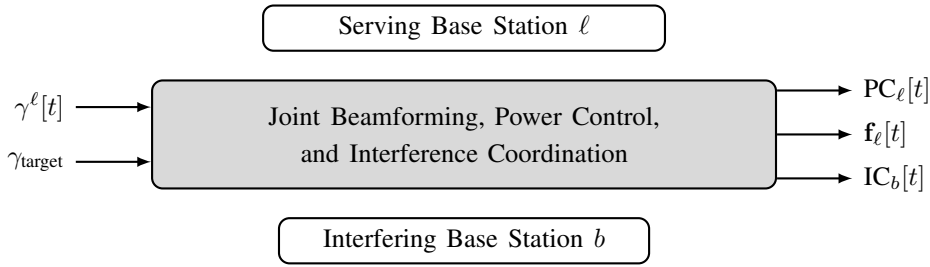
\begin{figure}[!t]
\centering
\linespread{1}
\begin{tikzpicture}[style=thick]
    \node [rectangle, draw, rounded corners, 
		text width=12em, text centered, minimum height=1em] at (0,0) (bs) {\small Serving Base Station $\ell$};
	
	\node [rectangle, draw, rounded corners, 
		text width=18em, text centered, minimum height=1em, inner sep=10pt, fill=gray!30, below=of bs, yshift=1.5em] (pc) {\small Joint Beamforming, Power Control, \\ and Interference Coordination};
	\node [rectangle, draw, rounded corners, 
		text width=11em, text centered, minimum height=1em,below=of pc,yshift=1.5em] (ibs) {\small Interfering Base Station $b$};

	\path [draw, latex-] (pc.175) -- ++(-1,0) node[left] {\small $\gamma^{\ell}[t]$};
	\path [draw, latex-] (pc.185) -- ++(-1,0) node[left] {\small $\gamma_\text{target}$};

	\path [draw, -latex] (pc.8) -- ++(1,0) node[text width=5em] [right] {\small PC$_\ell[t]$};
	\path [draw, -latex] (pc.0) -- ++(1,0) node[text width=5em] [right] {\small $\bff_\ell[t]$};
	\path [draw, -latex] (pc.352) -- ++(1,0) node[text width=5em] [right] {\small IC$_b[t]$};
\end{tikzpicture}%
\caption{Downlink joint beamforming, power control, and interference coordination module.} 
\label{fig:pc_module}
\end{figure}

\subsection{Brute Force}
The brute force PCIC algorithm uses an exhaustive search in the Euclidean space $\mathcal{P}$ per BS to optimize the SINR.  This algorithm solves \eqref{eq:optimization} and is the upper limit of the performance for jointly optimizing the SINR for the voice bearers in our problem.

\section{Deep Reinforcement Learning in mmWave Beamforming Power Control and Interference Coordination} \label{sec:bf_pcic}

\renewcommand{\thefootnote}{\fnsymbol{footnote}}

In this section, we present our proposed algorithms and quantify the changes in the SINR as a result of the movement of the UEs and optimization actions of the RL-based algorithm.

\begin{algorithm}[!t]
\small
\caption{\small Deep Reinforcement Learning in Joint Beamforming and PCIC (JB-PCIC)}
\label{alg:algorithm}
\DontPrintSemicolon
\KwIn{The downlink received SINR measured by the UEs.}
\KwOut{Sequence of beamforming, power control, and interference coordination commands to solve \eqref{eq:optimization}.}
  Initialize time, states, actions, and replay buffer $\mathcal{D}$.\;
  \Repeat {\text{\rm convergence or aborted}} {
  \Repeat {$t \ge T$} {
  $t := t  + 1$\;
  Observe current state $s_t$.\;
  $\epsilon := \max(\epsilon\cdot d, \epsilon_\text{min})$ \;
    Sample $r \sim \text{Uniform}(0,1)$\;
    \review{
  \eIf {$r \le \epsilon$} {
  Select an action $a_t \in \mathcal{A}$ at random.\;
  } {
  Select an action $a_t = \arg\max_{a^\prime} Q_\pi(s_t,a^\prime;\boldsymbol{\theta}_t)$. \;
  }
  }

  Compute $\gamma_\text{eff}^{\ell}[t]$ and $r_{s,s^\prime,a}[t;q]$ from \eqref{eq:rewards}.\;
  \If{$\gamma_\text{\rm eff}^{\ell}[t] < \gamma_\text{\rm min}$} {
  $r_{s,s^\prime,a}[t;q] := r_\text{min}$\;
  Abort episode.\;
  }
  
  Observe next state $s^\prime$.\;
Store experience $e[t] \triangleq (s_t, a_t, r_{s,s^\prime,a}, s^\prime)$ in $\mathcal{D}$.\;
Minibatch sample from $\mathcal{D}$ for experience $e_j\triangleq (s_j, a_j, r_j, s_{j+1})$.\; 

Set $y_j := r_j + \gamma\max_{a^\prime} Q_\pi(s_{j+1}, a^\prime; \boldsymbol{\theta}_t)$ \;

Perform SGD on $(y_j - Q_\pi(s_j, a_j; \boldsymbol{\theta}_t))^2$ to find $\boldsymbol{\theta}^\star$\;
Update $\boldsymbol{\theta}_t := \boldsymbol{\theta}^\star$ in the DQN and record loss $L_t$\;
$s_t := s^\prime$\;
 
}
}
\lIf{$\gamma_\text{\rm eff}^{\ell}[t] \ge \gamma_\text{\rm target}$} {$r_{s,s^\prime,a}[t;q] := r_{s,s^\prime,a}[t;q] + r_\text{max}$}
\end{algorithm}

\subsection{Proposed Algorithm}
We propose a DRL-based algorithm where the beamforming vectors and transmit powers at the base stations are jointly controlled to maximize the objective function in \eqref{eq:optimization}.  \reviewthird{The use of a string of bits as an action register enables us to jointly perform several actions concurrently.}

First, selecting the beamforming vector is performed as follows. The agent steps up or down the beamforming codebook using circular increments $(n + 1)$ or decrements $(n - 1)$
\begin{equation}
    n\mapsto \mathbf{f}_n[t] \colon n := (n \pm 1) \bmod M
    \label{eq:bf_stepping}
\end{equation}
for BSs $b$ and $\ell$ independently.  We monitor the change in $\gamma^\ell$ as a result of the change in the beamforming vector. \reviewsecond{We use a code gain of unity in computing $\gamma_\text{eff}^\ell$ for the data bearers (i.e., $\gamma_\text{eff}^\ell = \gamma^\ell$)}.

When the beamforming vectors are selected for a given UE, the agent also performs power control of that beam by changing the transmit power of the BS to this UE (or the interference coordination of other BSs).  The selection of the transmit power is governed by \eqref{eq:pc} and \eqref{eq:ic}, both of which define the set $\mathcal{P}$.

For proposed algorithm, the run time of the deep reinforcement learning is significantly faster than the upper bound algorithm for all antenna sizes $M$ as we show in Section~\ref{subsec:sim_results2}.  \review{Also, the reporting of the UE coordinates (i.e., longitude and latitude) to the BS instead of the channel state information reduces the reporting overhead from $M$ complex-valued elements to the two real-valued coordinates and its received SINR only.  If we assume that the reporting overhead for $M$ complex-valued elements is $2M$, then for reporting the UE coordinates, we achieve an overhead reduction gain of $1 - 1 / M$.}

We call our algorithm the \textit{joint beamforming, power control, and interference coordination} (JB-PCIC) algorithm.

\subsection{Brute Force}\label{subsec:optimal_solution}
The brute force beamforming and PCIC algorithm uses an exhaustive search in the Euclidean space $\mathcal{P}\times\mathcal{F}$ per BS to optimize the SINR.  As in the voice bearers brute force algorithm, this is also the upper limit in the performance for jointly optimizing the SINR in our problem.  While the size of $\mathcal{P}$ can be selected independently of the number of the antennas in the ULA  $M$, the size of $\mathcal{F}$ is directly related to $M$.  Similar to the brute force algorithm for voice bearers, this algorithm solves \eqref{eq:optimization} and may perform well for small $M$ and small number of BSs $L$ for data bearers.  However, we observe that with large $M$ the search time becomes prohibitive.  This is because the run time for this algorithm in $\mathcal{O}((\vert \mathcal{P}\vert \vert\mathcal{F}\vert)^L) = \mathcal{O}(M^L)$, which is much larger than the run time for the proposed algorithm, as we show in Section~\ref{subsec:sim_results2}.

\begin{table*}[!t]
\setlength\doublerulesep{0.5pt}
\caption{Reinforcement Learning Hyperparameters}
\label{table:rl_hyperparameters}
\centering
\vspace*{-1em}
\begin{tabular}{ lr lr } 
\hhline{====}
Parameter & Value & Parameter & Value \\
\hline
Discount factor $\gamma$ & 0.995 & Exploration rate decay $d$ &  0.9995 \\
Initial exploration rate $\epsilon$ & 1.000 & Minimum exploration rate $(\epsilon_\text{min}^\textrm{voice},\epsilon_\text{min}^\textrm{bf})$ & (0.15,0.10) \\
Number of states $\vert\mathcal{S}\vert$ & 8 & Number of actions $\vert\mathcal{A}\vert$  & 16  \\
Deep $Q$-Network width $H$ & 24 &Deep $Q$-Network depth & 2\\
\hhline{====}
\end{tabular}
\end{table*}

\begin{table*}[!t]
\setlength\doublerulesep{0.5pt}
\caption{Joint Beamforming Power Control Algorithm -- Radio Environment Parameters}
\label{table:rf_simulation}
\vspace*{-2em}
\begin{adjustwidth}{-1cm}{0cm}%
\centering
\begin{tabular}{ lrlr } 
\hhline{====}
Parameter & Value & Parameter & Value \\
 \hline
Base station (BS) maximum transmit power  $P_\text{BS}^{\rm max}$ & 46 dBm & Downlink frequency band & (2100 MHz, 28 GHz) \\
Cellular geometry & circular  & Cell radius $r$ & (350, 150) m \\
Propagation model (voice, bf) & (COST231,\cite{7522613}) & User equipment (UE) antenna gain & 0 dBi \\  
Antenna gain $(G_\text{TX}^\textrm{voice}, G_\text{TX}^\textrm{bf})$ & (11, 3) dBi & Inter-site distance $R$ & (525, 225) m \\ 
Max. number of UEs per BS $N$ & 10 & Number of multipaths $N_p$ & (15, 4) \\
Probability of LOS $p_\text{LOS}^\textrm{voice},p_\text{LOS}^\textrm{bf}$ & (0.9, 0.8) & UE average movement speed $v$ & (5, 2) km/h \\
Number of transmit antennas $M^\textrm{voice}, M^\textrm{bf}$ & (1,$\{4,8,16,32,64\}$) & Radio frame duration $T^\textrm{voice}, T^\textrm{bf}$ & (20, 10) ms \\
\hhline{====}
\end{tabular}
\end{adjustwidth}
\end{table*}

\section{Performance Measures}\label{sec:measures}
In this section we introduce the performance measures we use to benchmark our algorithms.

\subsection{Convergence} 
We define convergence $\zeta$ in terms of the episode at which the target SINR is fulfilled over the entire duration of $T$ for all UEs in the network.  We expect that as the number of antennas in the ULA $M$ increase, the convergence time $\zeta$ will also increase.  In voice, convergence as a function of $M$ is not applicable, since we only use single antennas.  For several random seeds, we take the aggregated percentile convergence episode.

\vspace*{-1em}
\subsection{Run time}
\reviewthird{While calculating the upper bound of the brute force algorithm run-time complexity is possible, obtaining a similar expression for the proposed deep $Q$-learning algorithm may be challenging due to lack of convergence and stability guarantees \cite{mnih2013playing}.  Therefore, we obtain the run time from simulation per antenna size $M$.}

\vspace*{-1em}
\subsection{Coverage} 
We build a {complement cumulative distribution function} (CCDF) of $\gamma_\text{eff}^\ell$ following \cite{6932503} by running the simulation many times and changing the random seed, effectively changing the way the users are dropped in the network.

\vspace*{-1em}
\subsection{Sum-rate capacity} 
Using the effective SINRs, we compute the average sum-rate capacity as%
\begin{equation}
    C = \frac{1}{T} \sum_{t=1}^T \sum_{j\in\{\ell,b\}} \log_2 ( 1 + \gamma_\text{eff}^j[t] )
    \label{eq:sum_rate}
\end{equation}
which is an indication of the data rate served by the network.  We then obtain the maximum sum-rate capacity resulting from computing \eqref{eq:sum_rate} over many episodes.

\section{Simulation Results}\label{sec:jbpcic_results}
In this section, we evaluate the performance of our RL-based proposed solutions in terms of the performance measures in Section \ref{sec:measures}. First, we describe the adopted setup in Section \ref{subsec:setup} before delving into the simulation results in Sections \ref{subsec:sim_results1} and \ref{subsec:sim_results2}.

\vspace*{-1em}
\subsection{Setup} \label{subsec:setup}

We adopt the network, signal, and channel models in Section \ref{sec:system}. 
The users in the urban cellular environment are uniformly distributed in its coverage area.   The users are moving at a speed $v$ with both log-normal shadow fading and small-scale fading.  The cell radius is $r$ and the inter-site distance $R = 1.5r$.  \review{For the voice bearer, we set the adaptive code rate $\beta$ between 1:3 to 1:1 based on reported SINR and use an AMR voice codec bitrate of 23.85 kbps and a voice activity factor $\nu = 0.8$}.  The users experience a probability of line of sight of $p_\text{LOS}$.  The rest of the parameters are shown in Table~\ref{table:rf_simulation}.
We set the target effective SINRs as:
\begin{align}
\nonumber \gamma_\text{target}^\text{voice} := 3 \,\text{dB},  \\
\gamma_\text{target}^\text{bf} := \gamma_0^\text{bf} + 10\log M \,\text{dB}
\label{eq:sinr_targets}
\end{align}
where $\gamma_0^\text{bf}$ is a constant threshold (i.e., not dependent on the antenna size).  We set the minimum SINR at $-3$ dB below which the episode is declared aborted and the session is unable to continue (i.e., dropped).

The hyperparameters required to tune the RL-based model are shown in Table~\ref{table:rl_hyperparameters}.  We refer to our source code \cite{my_jbpcic_code} for further implementation details.  Further, we run Algorithm~\ref{alg:algorithm} on the cellular network with its parameters in Table~\ref{table:rf_simulation}.   The simulated states $\mathcal{S}$ are setup as:
\begin{align}
    \nonumber
    (s_t^0, s_t^1) := \text{UE}_\ell (x[t],y[t]), \qquad & (s_t^2, s_t^3) := \text{UE}_b (x[t],y[t]), \\
    \nonumber
    s_t^4 := P_{\text{TX},\ell}[t], \qquad & s_t^5 := P_{\text{TX},b}[t], \\
    \nonumber
    s_t^6 := \bff_n^\ell[t], \qquad & s_t^7 := \bff_n^b[t],
    \nonumber
\end{align}
where $(x,y)$ are the Cartesian coordinates (i.e., longitude and latitude) of the given UE.  

\begin{figure}[!t]
\centering
\begin{tikzpicture}


\draw[draw=black] (0,0) rectangle (3,0.5) node[above=6pt,midway] (a1) {\small $2$} node[midway] {$\text{IC}_\ell[t]$}; 
\draw[draw=black] (3,0) rectangle (6,0.5) node[above=6pt,midway] (b1) {\small $2$} node[midway] {$\text{PC}_b[t]$}; 

\draw[draw=black] (0,-1) rectangle (1.5,-1.5) node[above=6pt,midway] (a2) {\small $1$} node[midway] {$\mathbf{f}_n^b[t]$};
\draw[draw=black] (1.5,-1) rectangle (3,-1.5) node[above=6pt,midway] {\small $1$} node[midway] {$\mathbf{f}_n^\ell[t]$};
\draw[draw=black] (3,-1) rectangle (4.5,-1.5) node[above=6pt,midway] {\small $1$} node[midway] {$\text{IC}_\ell[t]$}; 
\draw[draw=black] (4.5,-1) rectangle (6,-1.5)  node[above=6pt,midway] (b2) {\small $1$} node[midway] {$\text{PC}_b[t]$}; 

\draw[latex-latex] ( $(a2.west) + (-0.5cm,-1.2cm)$ ) -- node[fill=white] {$a_t\in\mathcal{A}$}  ( $(b2.east) + (0.5cm,-1.2cm)$ );

\node[left=of a1, xshift=-0.75cm, yshift=-0.5cm] {$q = 0\colon$};
\node[left=of a2, yshift=-0.5cm] {$q = 1\colon$};
\end{tikzpicture}
\caption{Binary encoding of joint beamforming, power control, and interference coordination actions using a string of bits in a register $\mathbf{a}$ for different bearer types ($q = 0$ for voice bearers and $q = 1$ for data bearers).}\label{fig:register}
\end{figure}
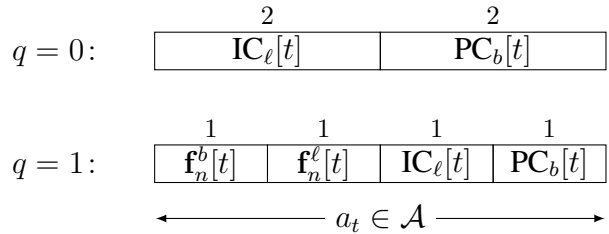

To derive the actions  $\mathcal{A}$, we exploit the fact that $\mathcal{F}$ and  $\mathcal{P}$  each has a cardinality that is a power of two.  This enables us to construct the binary encoding of the actions using a register $\mathbf{a}$ as shown in Fig.~\ref{fig:register}. \reviewsecond{With bitwise-AND, masks, and shifting, the joint beamforming, power control, and interference coordination commands can be derived.  We choose the following code: 

\begin{enumerate}
    \item When $q = 0$:
        \begin{itemize}
            \item $\mathbf{a}_{[0,1]} = 00$: decrease the transmit power of BS $b$ by $3$ dB.
            \item $\mathbf{a}_{[0,1]} = 01$: decrease the transmit power of BS $b$ by $1$ dB.
            \item $\mathbf{a}_{[0,1]}= 10$: increase the transmit power of BS $b$ by $1$ dB.
            \item $\mathbf{a}_{[0,1]} = 11$: increase the transmit power of BS $b$ by $3$ dB.
            \item $\mathbf{a}_{[2,3]}= 00$: decrease the transmit power of BS $\ell$ by $3$ dB.
            \item $\mathbf{a}_{[2,3]} = 01$: decrease the transmit power of BS $\ell$ by $1$ dB.
            \item $\mathbf{a}_{[2,3]}= 10$: increase the transmit power of BS $\ell$ by $1$ dB.
            \item $\mathbf{a}_{[2,3]}= 11$: increase the transmit power of BS $\ell$ by $3$ dB.
        \end{itemize}

    \item When $q = 1$:
        \begin{itemize}
            \item $\mathbf{a}_{[0]} = 0$: decrease the transmit power of BS $b$ by $1$ dB.
            \item $\mathbf{a}_{[0]} = 1$: increase the transmit power of BS $b$ by $1$ dB.
            \item $\mathbf{a}_{[1]} = 0$: decrease the transmit power of BS $\ell$ by $1$ dB.
            \item $\mathbf{a}_{[1]} = 1$: increase the transmit power of BS $\ell$ by $1$ dB.
            \item $\mathbf{a}_{[2]}= 0$: step down the beamforming\index{beamforming} codebook index of BS $\ell$.
            \item $\mathbf{a}_{[2]}= 1$: step up the beamforming\index{beamforming} codebook index of BS $\ell$.
            \item $\mathbf{a}_{[3]}= 0$: step down the beamforming\index{beamforming} codebook index of BS $b$.
            \item $\mathbf{a}_{[3]}= 1$: step up the beamforming\index{beamforming} codebook index of BS $b$.
        \end{itemize}%
\end{enumerate}
} %

Here, we can infer that $\mathcal{P} = \{\pm 1, \pm 3\}$ dB offset from the transmit power. \review{The choice of these values is motivated by 1) aligning with industry standards \cite{3gpp36213} which choose integers for power increments and 2) maintaining the non-convexity of the problem formulation \eqref{eq:optimization} by keeping the constraints discrete.} The actions to increase and decrease BS transmit powers are implemented as in \eqref{eq:pc} and \eqref{eq:ic}.  We introduce $3$-dB power steps for voice only to compensate for not using beamforming, which is aligned with the industry standards of not having beamforming for packetized voice bearers \cite{3gpp36213}. 

The reward we use in our proposed algorithms is divided into two tiers: 1) based on the relevance of the action taken and 2) based on whether the target SINR has been met or the SINR falls below the minimum.  We start by defining a function $\text{p}(\cdot)$ which returns one of the elements of $\mathcal{P}$ based on the chosen code.  \reviewsecond{Here, $\text{p}(00) = -3, \text{p}(01) = -1, \text{p}(10) = 1, \text{p}(11) = 3$.  Next, we write the received SINR due to the aforementioned encoded actions in $\mathbf{a}$ as $\gamma^b_{\mathbf{a}_{[0]}, \mathbf{a}_{[3]}}$ and $\gamma^\ell_{\mathbf{a}_{[1]}, \mathbf{a}_{[2]}}$ for BSs $b$ and $\ell$, respectively.

\review{We further write the joint reward for both voice and data bearers as follows:
\begin{equation}
\begin{aligned}
    r_{s,s^\prime,a}[t; q] & := \bigg (\text{p}(\mathbf{a}_{[0,1]}[t]) - \text{p}(\mathbf{a}_{[2,3]}[t])  \bigg ) (1 - q) + \bigg ( \gamma^b_{\mathbf{a}_{[0]}[t], \mathbf{a}_{[3]}[t]}  + \gamma^\ell_{\mathbf{a}_{[1]}[t], \mathbf{a}_{[2]}[t]}
     \bigg )  q
\end{aligned}
\label{eq:rewards}
\end{equation}
where $q = 0$ for voice bearers and $1$ for data bearers.}  We reward the agent the most per time step when a joint power control and beamforming action is taken for data bearers and when a joint power control and interference coordination takes place for a voice bearer.} We abort the episode if any of the constraints in \eqref{eq:optimization} becomes inactive.  At this stage, the RL agent receives a reward $r_{s,s^\prime,a}[t;q] := r_\text{min}$.  Either a penalty $r_\text{min}$ or a maximum reward $r_\text{max}$ is added based on whether the minimum $\gamma_\text{min}$ has been violated or $\gamma_\text{target}$ has been achieved as shown in Algorithm~\ref{alg:algorithm}.  Here, it is also clear that for data bearers the agent is rewarded more for searching in the beamforming codebook than attempting to power up or down.  However, for voice bearers, we reward the agent more if it chooses to power control the serving BS $b$ than if it chooses to control the interference from the other BS $\ell$.

In our simulations, we use a minibatch sample size of $N_\text{mb} = 32$ training examples. \reviewsecond{With $\vert\mathcal{A}\vert = 16$, the width of the DQN can be found using \cite{1189626} to be $H = \sqrt{(\vert\mathcal{A}\vert + 2)N_\text{mb}} = 24$.}  We refer to our code \cite{my_jbpcic_code} for details.

\subsection{Outcomes} \label{subsec:sim_results1}
\begin{enumerate}
\item Convergence: we study the normalized convergence under \eqref{eq:sinr_targets} where $\gamma_0^\text{bf} = 5$ dB. Every time step in an episode is equal to one radio subframe, the duration of which is 1 ms \cite{3gpp38211}. During this time the UE is likely to be using a sub-optimal selection of beam obtained from a prior iteration.  This would cause the UE throughput to degrade by a factor as we show in Section~\ref{subsec:sim_results2}.  As the size of the ULA $M$ increases, the number of episodes required converge increases with minimal effect of the  constant threshold $\gamma_0^\text{bf}$ since $M \gg \gamma_0^\text{bf}$.  This is justified since the number of attempts to traverse the beamforming codebook increases almost linearly with the increase of $M$.  

\item Run time: we study the normalized run time and observe that as the number of antennas $M$ increase, so does the run-time complexity for the proposed algorithm.  This is justified due to the increase in the number of beams required for the algorithm to search through to increase the joint SINR.

\item Coverage: for voice bearers we observe that the coverage as defined by the SINR CCDF improves everywhere.  For data bearers, the coverage improves where the SINR monotonically increases with the increase in $M$ which is expected because the beamforming array gain increases with an increase in $M$.

\item Sum-rate: the sum-rate capacity increases logarithmically as a result of the increase of $M$, which is justified using \eqref{eq:sinr_final} and \eqref{eq:sum_rate}.
\end{enumerate}


\subsection{Figures} \label{subsec:sim_results2}


\begin{figure}[!t]
\centering 
{\resizebox{0.75\textwidth}{!}{
\begin{tikzpicture}

\definecolor{color0}{rgb}{0.12156862745098,0.466666666666667,0.705882352941177}
\definecolor{color1}{rgb}{1,0.498039215686275,0.0549019607843137}
\definecolor{color2}{rgb}{0.172549019607843,0.627450980392157,0.172549019607843}
\definecolor{color3}{rgb}{0.83921568627451,0.152941176470588,0.156862745098039}

\begin{axis}[
width=4.0in,
height=3.0in,
legend cell align={left},
legend entries={{Deep $Q$-learning (proposed)}, {Tabular $Q$-learning}, {Fixed Power Allocation (FPA)}, {Brute Force}},
legend style={at={(0.01,0.005)}, anchor=south west, draw=white!80.0!black, nodes={scale=0.7, transform shape}},
tick align=outside,
tick pos=left,
x grid style={white!69.01960784313725!black, dashed},
xlabel={Voice effective SINR, $\gamma_\text{eff}^\text{voice}$ [dB]},
xmajorgrids,
xmin=0, xmax=15,
y grid style={white!69.01960784313725!black, dashed},
ylabel={CCDF},
ymajorgrids,
ymin=-0, ymax=1, 
ytick={-0.2,0,0.2,0.4,0.6,0.8,1,1.2},
yticklabels={$-0.2$,$0.0$,$0.2$,$0.4$,$0.6$,$0.8$,$1.0$,$1.2$}
]

\addplot [line width=1.2pt, color1, mark=square*, mark size=2, mark repeat=10, mark options={solid}]
table [row sep=\\]{%
4.73057816718125	1 \\
5.10169295950839	0.975 \\
5.28725035567197	0.975 \\
5.47280775183554	0.975 \\
5.65836514799911	0.975 \\
5.84392254416268	0.975 \\
6.02947994032625	0.95 \\
6.21503733648983	0.95 \\
6.4005947326534	0.95 \\
6.58615212881697	0.95 \\
6.77170952498054	0.95 \\
6.95726692114411	0.95 \\
7.14282431730768	0.95 \\
7.32838171347126	0.95 \\
7.51393910963483	0.925 \\
7.6994965057984	0.925 \\
7.88505390196197	0.85 \\
8.07061129812554	0.85 \\
8.25616869428912	0.85 \\
8.44172609045269	0.8 \\
8.62728348661626	0.8 \\
8.81284088277983	0.775 \\
8.9983982789434	0.725 \\
9.18395567510697	0.675 \\
9.36951307127055	0.575 \\
9.55507046743412	0.575 \\
9.74062786359769	0.525 \\
9.92618525976126	0.425 \\
10.1117426559248	0.425 \\
10.2973000520884	0.350000000000001 \\
10.482857448252	0.325000000000001 \\
10.6684148444155	0.2 \\
10.8539722405791	0.2 \\
11.0395296367427	0.15 \\
11.2250870329063	0.0999999999999999 \\
11.4106444290698	0.075 \\
11.5962018252334	0.075 \\
11.781759221397	0.075 \\
11.9673166175606	0.075 \\
12.1528740137241	0.0499999999999999 \\
12.3384314098877	0.0499999999999999 \\
12.5239888060513	0.0499999999999999 \\
12.7095462022148	0.025 \\
12.8951035983784	0.025 \\
13.080660994542	0.025 \\
13.2662183907056	0.025 \\
13.4517757868691	0.025 \\
13.6373331830327	0.025 \\
13.8228905791963	0.025 \\
14.0084479753598	0.025 \\
14.1940053715234	-2.22044604925031e-16 \\
};
\addplot [line width=1.2pt, color2, mark=*, mark size=2, mark repeat=10, mark options={solid}]
table [row sep=\\]{%
-0.104306296327642	1 \\
0.448797315122734	0.975 \\
0.725349120847922	0.975 \\
1.00190092657311	0.975 \\
1.2784527322983	0.975 \\
1.55500453802349	0.975 \\
1.83155634374868	0.975 \\
2.10810814947386	0.95 \\
2.38465995519905	0.95 \\
2.66121176092424	0.95 \\
2.93776356664943	0.925 \\
3.21431537237462	0.8 \\
3.4908671780998	0.8 \\
3.76741898382499	0.8 \\
4.04397078955018	0.8 \\
4.32052259527537	0.775 \\
4.59707440100056	0.775 \\
4.87362620672575	0.775 \\
5.15017801245093	0.775 \\
5.42672981817612	0.75 \\
5.70328162390131	0.75 \\
5.9798334296265	0.6 \\
6.25638523535169	0.55 \\
6.53293704107687	0.55 \\
6.80948884680206	0.55 \\
7.08604065252725	0.525 \\
7.36259245825244	0.525 \\
7.63914426397763	0.525 \\
7.91569606970281	0.525 \\
8.192247875428	0.5 \\
8.46879968115319	0.5 \\
8.74535148687838	0.475 \\
9.02190329260357	0.224999999999999 \\
9.29845509832876	0.199999999999999 \\
9.57500690405394	0.174999999999999 \\
9.85155870977913	0.174999999999999 \\
10.1281105155043	0.149999999999999 \\
10.4046623212295	0.149999999999999 \\
10.6812141269547	0.124999999999999 \\
10.9577659326799	0.0249999999999998 \\
11.2343177384051	0.0249999999999998 \\
11.5108695441303	0.0249999999999998 \\
11.7874213498554	0.0249999999999998 \\
12.0639731555806	0.0249999999999998 \\
12.3405249613058	0.0249999999999998 \\
12.617076767031	0.0249999999999998 \\
12.8936285727562	0.0249999999999998 \\
13.1701803784814	0.0249999999999998 \\
13.4467321842066	0.0249999999999998 \\
13.7232839899318	0.0249999999999998 \\
13.999835795657	-2.22044604925031e-16 \\
};
\addplot [line width=1.2pt, color3, mark=*, mark size=2, mark repeat=10, mark options={solid}]
table [row sep=\\]{%
0.178764810619244	1 \\
0.81738016516405	0.7625 \\
1.03025195001232	0.7625 \\
1.24312373486059	0.7625 \\
1.45599551970885	0.7625 \\
1.66886730455712	0.7625 \\
1.88173908940539	0.7625 \\
2.09461087425366	0.7625 \\
2.30748265910193	0.7625 \\
2.5203544439502	0.7625 \\
2.73322622879847	0.7625 \\
2.94609801364673	0.625 \\
3.158969798495	0.525 \\
3.37184158334327	0.525 \\
3.79758515303981	0.525 \\
4.01045693788808	0.525 \\
4.22332872273634	0.525 \\
4.43620050758461	0.525 \\
4.64907229243288	0.525 \\
4.86194407728115	0.525 \\
5.07481586212942	0.525 \\
5.28768764697769	0.500000000000001 \\
5.50055943182596	0.500000000000001 \\
5.71343121667422	0.500000000000001 \\
5.92630300152249	0.500000000000001 \\
6.13917478637076	0.500000000000001 \\
6.35204657121903	0.500000000000001 \\
6.5649183560673	0.500000000000001 \\
6.77779014091557	0.500000000000001 \\
6.99066192576383	0.4875 \\
7.2035337106121	0.4875 \\
7.41640549546037	0.4875 \\
7.62927728030864	0.4875 \\
7.84214906515691	0.475000000000001 \\
8.05502085000518	0.375000000000001 \\
8.26789263485345	0.300000000000001 \\
8.48076441970171	0.275000000000001 \\
8.69363620454998	0.275000000000001 \\
8.90650798939825	0.262500000000001 \\
9.11937977424652	0.262500000000001 \\
9.33225155909479	0.262500000000001 \\
9.54512334394306	0.262500000000001 \\
9.75799512879133	0.1 \\
9.97086691363959	0.0125000000000002 \\
10.1837386984879	0.0125000000000002 \\
10.3966104833361	0.0125000000000002 \\
10.6094822681844	0.0125000000000002 \\
10.8223540530327	0.0125000000000002 \\
11.0352258378809	0 \\
};
\addplot [line width=1.2pt, color0, dashed, mark=triangle*, mark size=3, mark repeat=10, mark phase=5, mark options={solid}]
table [row sep=\\]{%
4.73057816718125	1 \\
5.10169295950839	0.975 \\
5.28725035567197	0.975 \\
5.47280775183554	0.975 \\
5.65836514799911	0.975 \\
5.84392254416268	0.975 \\
6.02947994032625	0.95 \\
6.21503733648983	0.95 \\
6.4005947326534	0.95 \\
6.58615212881697	0.95 \\
6.77170952498054	0.95 \\
6.95726692114411	0.95 \\
7.14282431730768	0.95 \\
7.32838171347126	0.95 \\
7.51393910963483	0.925 \\
7.6994965057984	0.925 \\
7.88505390196197	0.85 \\
8.07061129812554	0.85 \\
8.25616869428912	0.85 \\
8.44172609045269	0.8 \\
8.62728348661626	0.8 \\
8.81284088277983	0.775 \\
8.9983982789434	0.725 \\
9.18395567510697	0.675 \\
9.36951307127055	0.575 \\
9.55507046743412	0.575 \\
9.74062786359769	0.525 \\
9.92618525976126	0.425 \\
10.1117426559248	0.425 \\
10.2973000520884	0.350000000000001 \\
10.482857448252	0.325000000000001 \\
10.6684148444155	0.2 \\
10.8539722405791	0.2 \\
11.0395296367427	0.15 \\
11.2250870329063	0.0999999999999999 \\
11.4106444290698	0.075 \\
11.5962018252334	0.075 \\
11.781759221397	0.075 \\
11.9673166175606	0.075 \\
12.1528740137241	0.0499999999999999 \\
12.3384314098877	0.0499999999999999 \\
12.5239888060513	0.0499999999999999 \\
12.7095462022148	0.025 \\
12.8951035983784	0.025 \\
13.080660994542	0.025 \\
13.2662183907056	0.025 \\
13.4517757868691	0.025 \\
13.6373331830327	0.025 \\
13.8228905791963	0.025 \\
14.0084479753598	0.025 \\
14.1940053715234	-2.22044604925031e-16 \\
};




\end{axis}

\end{tikzpicture}}}%
\vspace*{-1em}
\caption{Coverage CCDF plot of $\gamma_\text{eff}^\textrm{voice}$ for three different voice power control and interference coordination algorithms.}
\label{fig:voice_ccdf}
\end{figure}
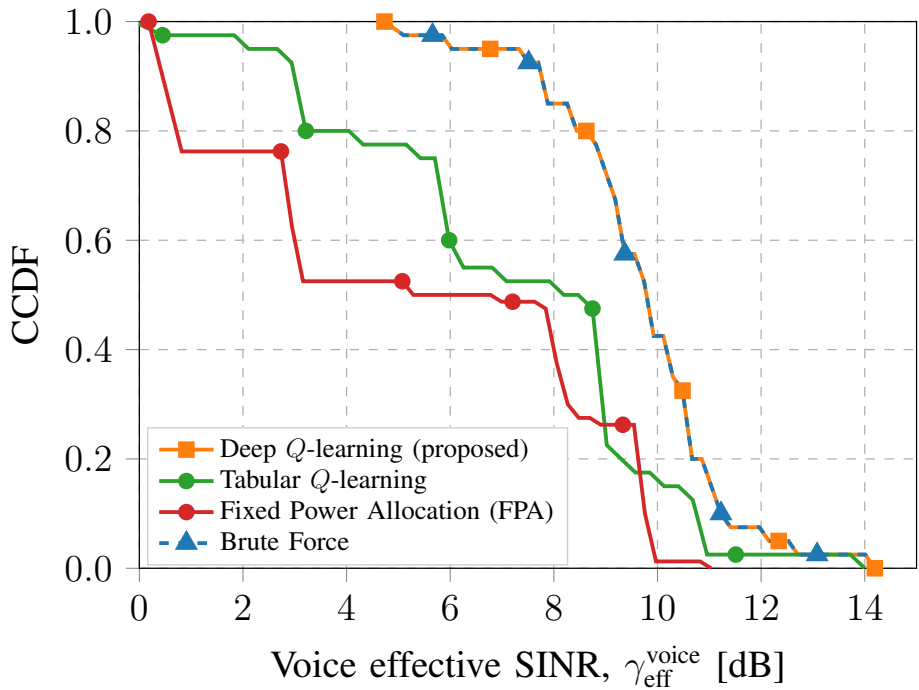

\reviewthird{
Fig.~\ref{fig:voice_ccdf} shows the CCDF of the effective SINR $\gamma_\text{eff}$ for the voice PCIC algorithms all for the same episode.  This episode generates the highest reward.  Here we see that the FPA algorithm has the worst performance especially at the cell edge (i.e., low effective SINR regime), which is expected since FPA has no power control or interference coordination.
The tabular implementation of our proposed algorithm has better performance compared with the FPA.  This is since power control and interference coordination are introduced to the base stations, though not as effectively, which explains why close to $\gamma_\text{eff} = 9$ dB tabular $Q$-learning PCIC underperforms FPA.
Further, we observe that deep $Q$-learning outperforms the tabular $Q$-learning implementation of the PCIC algorithm, since deep $Q$-learning has resulted in a higher reward compared to tabular $Q$-learning.  This is because deep $Q$-learning has converged at a better solution (identical to the solution obtained through brute force), unlike the tabular $Q$-learning the convergence of which may have been impeded by the choice of a initialization of the state-action value function.
However, as the effective SINR $\gamma_\text{eff}$ approaches $13$ dB, the users are close to the BS center and therefore all power control algorithms perform almost similarly thereafter.
}
\begin{figure}[!t] 
\centering 
{\resizebox{0.75\textwidth}{!}{
\begin{tikzpicture}

\definecolor{color0}{rgb}{0.12156862745098,0.466666666666667,0.705882352941177}
\definecolor{color1}{rgb}{1,0.498039215686275,0.0549019607843137}
\definecolor{color2}{rgb}{0.172549019607843,0.627450980392157,0.172549019607843}
\definecolor{color3}{rgb}{0.83921568627451,0.152941176470588,0.156862745098039}
\definecolor{color4}{rgb}{0.580392156862745,0.403921568627451,0.741176470588235}

\begin{axis}[
width=3.6in,
height=2.8in,
legend cell align={left},
legend entries={{$M = 4$},{$M = 8$},{$M = 16$},{$M = 32$},{$M = 64$}},
legend style={at={(0.72,0.62)}, anchor=south west, draw=white!80.0!black, nodes={scale=0.75, transform shape}},
tick align=outside,
tick pos=left,
x grid style={white!69.01960784313725!black, dashed},
xlabel={$\gamma_\text{eff}$ [dB]},
xmajorgrids,
xmin=0, xmax=70,
xtick={0,10,...,70},
y grid style={white!69.01960784313725!black, dashed},
ylabel={CCDF}, 
ymajorgrids,
ymin=0, ymax=1, 
ytick={0,0.1,...,1.1},
]
\addplot [color0, line width=1.2pt, mark=x, mark options={solid}]
table [row sep=\\]{%
-5.78984477785721	1 \\
-0.191956983251494	0.935294117647059 \\
2.60698691405136	0.875 \\
5.40593081135422	0.799509803921569 \\
8.20487470865708	0.715686274509804 \\
11.0038186059599	0.62156862745098 \\
13.8027625032628	0.511764705882353 \\
16.6017064005657	0.396078431372549 \\
19.4006502978685	0.268137254901961 \\
22.1995941951714	0.132352941176471 \\
24.9985380924742	0 \\
};
\addplot [color1, line width=1.2pt, mark=diamond*, mark options={solid}]
table [row sep=\\]{%
-6.7600923410502	1 \\
0.829146617541442	0.934065934065934 \\
4.62376609683726	0.877289377289377 \\
8.41838557613308	0.802808302808303 \\
12.2130050554289	0.715506715506716 \\
16.0076245347247	0.605616605616606 \\
19.8022440140205	0.492673992673993 \\
23.5968634933164	0.35958485958486 \\
27.3914829726122	0.235042735042735 \\
31.186102451908	0.104395604395604 \\
34.9807219312038	0 \\
};
\addplot [color2, line width=1.2pt, mark=square*, mark options={solid}]
table [row sep=\\]{%
-7.75110439520845	1 \\
1.83725120512764	0.931034482758621 \\
6.63142900529568	0.857471264367816 \\
11.4256068054637	0.768582375478927 \\
16.2197846056318	0.672030651340996 \\
21.0139624057998	0.545593869731801 \\
25.8081402059679	0.419157088122605 \\
30.6023180061359	0.311111111111111 \\
35.396495806304	0.196168582375479 \\
40.190673606472	0.096551724137931 \\
44.98485140664	0 \\
};
\addplot [color3, line width=1.2pt, mark=triangle*, mark options={solid}]
table [row sep=\\]{%
-8.73441649678819	1 \\
2.84771381734341	0.929595827900913 \\
8.63877897440921	0.852672750977836 \\
14.429844131475	0.753585397653194 \\
20.2209092885408	0.641460234680574 \\
26.0119744456066	0.509778357235984 \\
31.8030396026724	0.36245110821382 \\
37.5941047597382	0.224250325945241 \\
43.385169916804	0.135593220338983 \\
49.1762350738698	0.0664928292046937 \\
54.9673002309356	0 \\
};
\addplot [color4, line width=1.2pt, mark=*, mark options={solid}]
table [row sep=\\]{%
-9.67331081889672	1 \\
3.88736962192217	0.910984848484849 \\
10.6677098423316	0.821969696969697 \\
17.4480500627411	0.723484848484848 \\
24.2283902831505	0.596590909090909 \\
31.00873050356	0.450757575757576 \\
37.7890707239694	0.335227272727273 \\
44.5694109443788	0.206439393939394 \\
51.3497511647883	0.102272727272727 \\
58.1300913851977	0.0606060606060604 \\
64.9104316056072	0 \\
};
\end{axis}

\end{tikzpicture}}}%
\caption{Coverage CCDF plot of the effective SINR $\gamma_\text{eff}$ for the proposed deep $Q$-learning algorithm vs. the number of antennas $M$.}
\label{fig:mmwave_ccdf}
\end{figure}
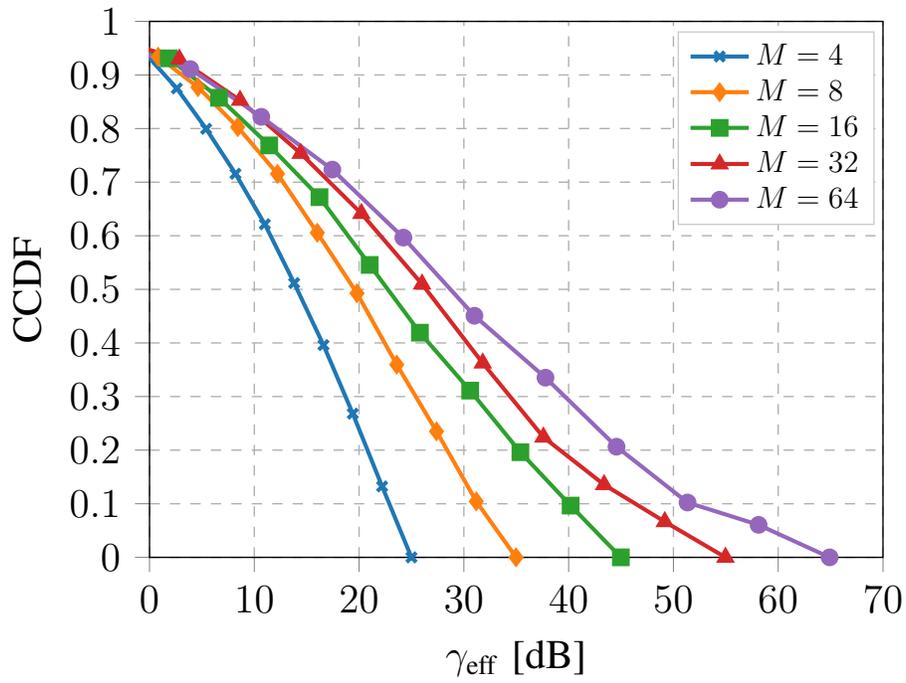

\begin{figure}[!t]
\centering 
{\resizebox{0.75\textwidth}{!}{
\begin{tikzpicture}

\begin{semilogyaxis}[
width=4.0in,
height=3.0in,
legend cell align={left},
legend entries={{JB-PCIC},{Upper Bound}},
legend style={at={(0.01,0.82)}, anchor=south west, draw=white!80.0!black, nodes={scale=0.8, transform shape}},
tick align=outside,
tick pos=left,
x grid style={white!69.01960784313725!black, dashed},
xlabel={$M$},
xmajorgrids,
xmin=0, xmax=64,
xtick={4,8,16,32,64},
y grid style={white!69.01960784313725!black, dashed},
ylabel={Normalized run time},
ymajorgrids,
ymin=1e-4, ymax=1,
]
\addplot [line width=1.2pt, black, mark=square*, mark size=2, mark options={solid}]
table [row sep=\\]{%
 4     0.000170413483830708 \\ 
 8     0.000172184645162987 \\ 
 16     0.000252488666278169 \\ 
 32     0.000300558753998562 \\ 
 64     0.000524748528913154 \\ 
};

\addplot [line width=1.2pt, red, mark=triangle*, mark size=2, mark options={solid}]
table [row sep=\\]{%
 4     0.00435529204096564 \\ 
 8     0.0153368251377943 \\ 
 16     0.0750825813587176 \\ 
 32     0.26626988941241 \\ 
 64     1 \\ 
};





\end{semilogyaxis}

\end{tikzpicture}}}%
\caption{The normalized run times for the proposed deep $Q$-learning algorithm as a function of the number of antennas $M$.}
\label{fig:runtimes}
\end{figure}
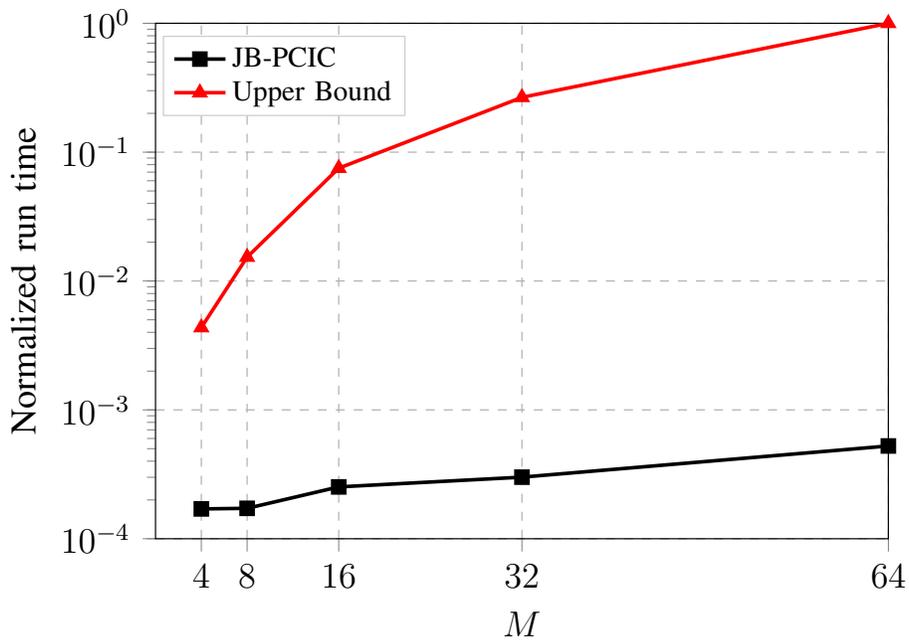

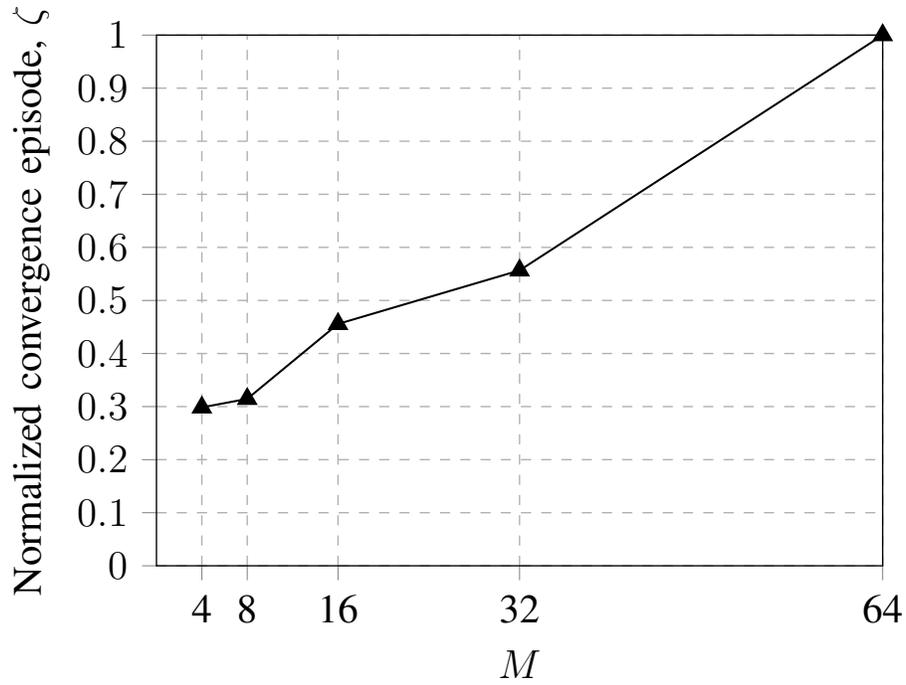
\begin{figure}[!t]
\centering 
{\resizebox{0.75\textwidth}{!}{
\begin{tikzpicture}

\begin{axis}[
width=3.6in,
height=2.8in,
legend cell align={left},
tick align=outside,
tick pos=left,
x grid style={white!69.01960784313725!black, dashed},
xlabel={$M$},
xmajorgrids,
xmin=0, xmax=64,
xtick={4,8,16,32,64},
xticklabels={4,8,16,32,64},
y grid style={white!69.01960784313725!black,dashed},
ylabel={Normalized convergence episode, $\zeta$},
ymajorgrids,
ymin=0, ymax=1,
ytick={0,0.1,...,1.1},
]
\addplot [semithick, black, mark=triangle*, mark size=3, mark options={solid}, forget plot]
table [row sep=\\]{%
4	0.298452012383901 \\
8	0.314551083591331 \\
16	0.455727554179567 \\
32	0.556656346749226 \\
64	1 \\
};





\end{axis}

\end{tikzpicture}}}
\caption{The normalized convergence time for the proposed deep $Q$-learning algorithm as a function of the number of antennas $M$.}
\label{fig:convergence_vs_M}
\end{figure}

We show the coverage CCDF in Fig.~\ref{fig:mmwave_ccdf}.  As $M$ increases, so does the probability of achieving a given effective SINR, since the effective SINR depends on the beamforming array gain which is a function of $M$ as stated earlier. \reviewthird{The improvement in the run time is shown in Fig.~\ref{fig:runtimes}. The brute force algorithm has a significantly larger run time compared to the proposed algorithm.  The run time increases as the number of antennas $M$ increase, though much steeper in the brute force algorithm, due to the exponential nature of the run-time complexity.   At $M = 4$, only 4\% of the run time of the brute force algorithm was needed for our proposed algorithm.} In Fig.~\ref{fig:convergence_vs_M}, \review{at smaller ULA sizes $M$, the impact of the constant threshold $\gamma_0^\text{bf}$ becomes dominant and it takes almost similar times to converge for values of $M$.  This is likely to be due to the wider beams in the grid of beams, which are able to cover the UEs moving at speeds $v$.  However, for the large antenna size regime, as the size of the ULA $M$ increases, the number of episodes required converge increases with minimal effect of $\gamma_0^\text{bf}$ as we explained earlier.}  This is due to the longer time required for the agent to search through a grid of beams of size $\vert\cF\vert$, which are typically narrower at large $M$.  This causes the agent to spend longer time to meet the target SINR.  \review{This time or delay is linear in $M$ as we expect based on Section~\ref{sec:bf_pcic} since $\vert\mathcal{F}\vert$ is linear in $M$.  This delay can have a negative impact on the throughput and voice frames of the data and voice bearers respectively.  If we assume the data bearer transmits $b$ bits over a total duration of $T^\text{bf}$ for beamformed data bearers, then the impact of the convergence time would cause these $b$ bits to be transmitted over a duration of $T^\text{bf}\zeta$.  The throughput due to convergence then becomes $b/T^\text{bf}\zeta$.  For voice, the number of lost voice frames due to this convergence time is $\lceil \nu \zeta \rceil$. }   

\begin{figure}[!t]
\centering 
{\resizebox{0.75\textwidth}{!}{
\begin{tikzpicture}

\begin{axis}[
width=3.6in,
height=2.8in,
tick align=outside,
tick pos=left,
every y tick label/.append style={blue},
x grid style={white!69.01960784313725!black, dashed},
xlabel={$M$},
xmin=0, xmax=64,
xtick={4,8,16,32,64},
y grid style={white!69.01960784313725!black, dashed},
ylabel={Normalized transmit power},
ymin=0, ymax=1.3,
ytick={0,0.2,...,1.2},
grid=both,
]

\addplot [line width=1.2pt, cyan, mark=triangle*, mark size=3, mark phase=1, mark repeat=2, mark options={solid}]
table [row sep=\\]{%
4	1 \\
8	1 \\
16	1 \\
32	1 \\
64	1 \\
};
\addplot [line width=1.2pt, blue, dashed, mark=*, mark size=3, mark phase=2, mark repeat=2, mark options={solid}]
table [row sep=\\]{%
4	1 \\
8	1 \\
16	1 \\
32	1 \\
64	1 \\
};




\end{axis}

\begin{axis}[
width=3.6in,
height=2.8in,
axis y line=right,
axis line style={-},
every axis label/.append style ={black},
every tick label/.append style={red},
legend cell align={left},
legend entries={{TX Power JB-PCIC},{TX Power Brute Force},{SINR JB-PCIC},{SINR Brute Force}},
legend style={at={(0.30,0.39)}, text=black, anchor=north west, draw=white!80.0!black},
xtick={4,8,16,32,64},
xticklabels={,,},
xmin=0, xmax=64,
ylabel={Achievable effective SINR, $\gamma_\text{eff}$ [dB]},
ymin=20, ymax=65,
ytick={20,30,40,...,70},
ytick pos=right,
]
 \addlegendimage{cyan,mark=triangle*}
 \addlegendimage{blue,mark=*}
 \addlegendimage{white!30.0!red,mark=triangle}
 \addlegendimage{black!25.0!red,mark=*}

\addplot [line width=1.2pt, white!30.0!red, mark=triangle, mark size=3, mark phase=1, mark repeat=2, mark options={solid}]
table [row sep=\\]{%
4	24.9805437813393 \\
8	34.9915904013779 \\
16	44.9667559508266 \\
32	54.773338631735 \\
64	64.7985759935161 \\
};
\addplot [line width=1.2pt, black!25.0!red, dashed, mark=*, mark size=3, mark phase=2, mark repeat=2, mark options={solid}]
table [row sep=\\]{%
4	24.9977288984133 \\
8	34.9971300736623 \\
16	44.9974758502695 \\
32	54.9307821934073 \\
64	64.9662106521746 \\
};




\end{axis}

\end{tikzpicture}}}%
\vspace*{-1em}
\caption{Achievable SINR and normalized transmit power for both the brute force and proposed JB-PCIC algorithms as a function of the number of antennas $M$.}
\label{fig:sinr_power_vs_M}
\end{figure}
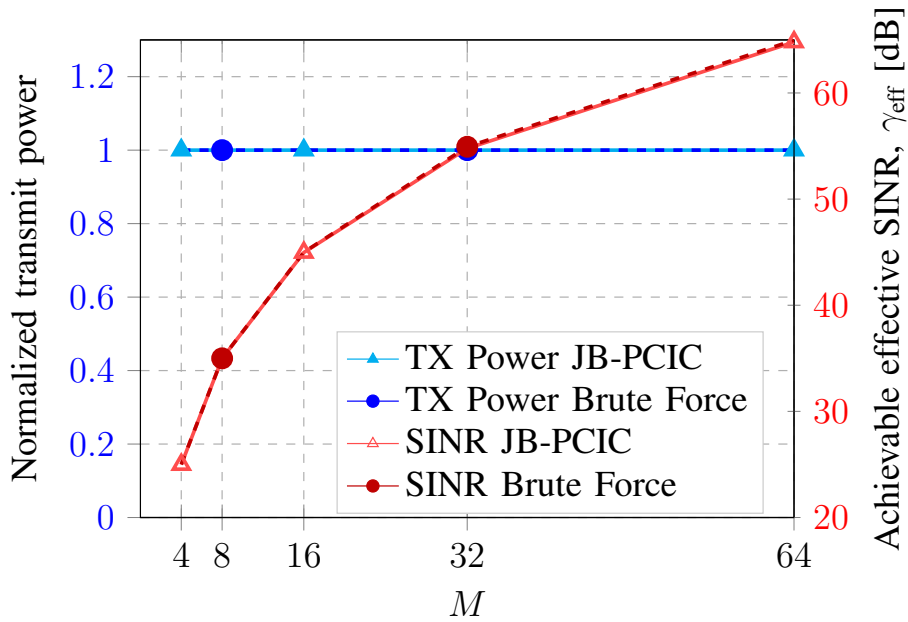

The achieved SINR is proportional to the ULA antenna size $M$ as shown in Fig.~\ref{fig:sinr_power_vs_M}.  This is expected as the beamforming array gain is $\Vert \mathbf{f}_b\Vert ^2 \le M$. The transmit power is almost equal to the maximum.
Fig.~\ref{fig:sinr_power_vs_M} also shows the relative performance of JB-PCIC compared with the  \review{brute force performance outlined in Section~\ref{subsec:optimal_solution}}.  We observe that the performance gap of both the transmit power of the base stations and the SINR is almost diminished all across $M$.  \reviewthird{ This is because of the DQN ability to estimate the function that leads to the upper limit of the performance.   Further, we observe that the solution for the race condition is for both BSs to transmit at maximum power.}

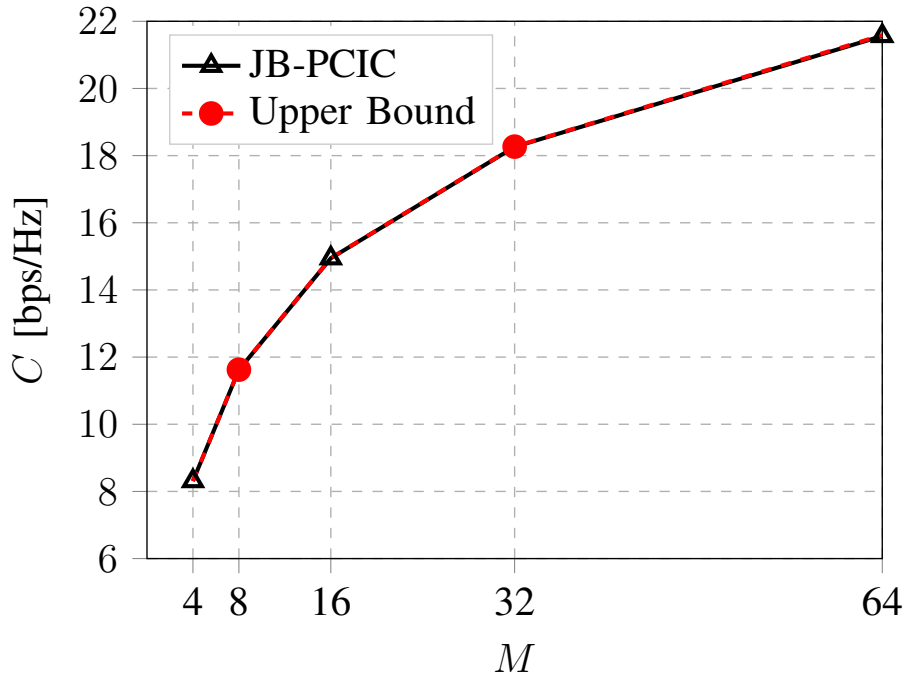
\begin{figure}[!t] 
\centering 
{\resizebox{0.75\textwidth}{!}{
\begin{tikzpicture}

\begin{axis}[
width=3.6in,
height=2.8in,
legend cell align={left},
legend entries={{JB-PCIC},{Upper Bound}},
legend style={at={(0.03,0.77)}, anchor=south west, draw=white!80.0!black},
tick align=outside,
tick pos=left,
x grid style={white!69.01960784313725!black, dashed},
xlabel={$M$},
xmajorgrids,
xmin=0, xmax=64,
xtick={4,8,16,32,64},
xticklabels={4,8,16,32,64},
y grid style={white!69.01960784313725!black, dashed},
ylabel={$C$ [bps/Hz]},
ymajorgrids,
ymin=6, ymax=22,
ytick={6,8,...,22},
]
\addplot [black, line width=1.2pt, mark=triangle, mark size=3, mark phase=1, mark repeat=2] 
table [row sep=\\]{%
4	8.30889113717368 \\
8	11.6208024733734 \\
16	14.9436899539134 \\
32	18.2597464904006 \\
64	21.56277910592 \\
};
\addplot [red, dashed, line width=1.2pt, mark=*, mark phase=2, mark repeat=2, mark size=3, mark options={solid}]
table [row sep=\\]{%
4	8.30806020370828 \\
8	11.6258280866722 \\
16	14.9480065176929 \\
32	18.2676149284105 \\
64	21.5882452296371 \\
};




\end{axis}

\end{tikzpicture}}}%
\vspace*{-1em}
\caption{Sum-rate capacity of the convergence episode as a function of the number of antennas $M$.}
\label{fig:sum_rate}
\end{figure}

Finally, Fig.~\ref{fig:sum_rate} shows the sum-rate capacity of both the JB-PCIC \review{algorithm} and the \review{upper limit of performance}. Similarly, the performance gap diminishes across all $M$ for the same reason discussed earlier.

\section{Conclusion}\label{sec:jbpcic_conclusion}

In this paper, we sought to maximize the downlink SINR in a multi-access OFDM cellular network from a multi-antenna base station to single-antenna user equipment.  The user equipment experienced interference from other multi-antenna base stations.  Our system used sub-6 GHz frequencies for voice and mmWave frequencies for data.  We assumed that each base station could select a beamforming vector from a finite set.  The power control commands were also from a finite set.  We showed that a closed-form solution did not exist, and that finding the optimum answer required an exhaustive search.  An exhaustive search had a run time exponential in the number of base stations.

To avoid an exhaustive search, we developed a joint beamforming, power control, and interference coordination algorithm (JB-PCIC) using deep reinforcement learning.  \review{This algorithm resides at a central location and receives UE measurements over the backhaul.} For voice bearers, our proposed algorithm outperformed both the tabular $Q$-learning algorithm and the industry standard fixed power allocation algorithm. 

Our proposed algorithm for joint beamforming, power control and interference coordinations requires that the UE sends its coordinates and its received SINR every millisecond to the base station.  The proposed algorithm, however, does not require the knowledge of the channel state information, which removes the need for channel estimation and the associated training sequences.  Moreover, the overall amount of feedback from the UE is reduced because the UE sends its coordinates and would not need to send explicit commands for beamforming vector changes, power control, or interference coordination.

\bibliographystyle{IEEEtran}
\bibliography{Reinforcement.bib,AlkhateebRef.bib}

\begin{thebibliography}{10}
\providecommand{\url}[1]{#1}
\csname url@samestyle\endcsname
\providecommand{\newblock}{\relax}
\providecommand{\bibinfo}[2]{#2}
\providecommand{\BIBentrySTDinterwordspacing}{\spaceskip=0pt\relax}
\providecommand{\BIBentryALTinterwordstretchfactor}{4}
\providecommand{\BIBentryALTinterwordspacing}{\spaceskip=\fontdimen2\font plus
\BIBentryALTinterwordstretchfactor\fontdimen3\font minus
  \fontdimen4\font\relax}
\providecommand{\BIBforeignlanguage}[2]{{%
\expandafter\ifx\csname l@#1\endcsname\relax
\typeout{** WARNING: IEEEtran.bst: No hyphenation pattern has been}%
\typeout{** loaded for the language `#1'. Using the pattern for}%
\typeout{** the default language instead.}%
\else
\language=\csname l@#1\endcsname
\fi
#2}}
\providecommand{\BIBdecl}{\relax}
\BIBdecl

\bibitem{8645168}
F.~B. {Mismar} and B.~L. {Evans}, ``{Q-Learning Algorithm for VoLTE Closed Loop
  Power Control in Indoor Small Cells},'' in \emph{Proc. Asilomar Conference on
  Signals, Systems, and Computers}, Oct. 2018.

\bibitem{5683371}
S.~{Yun} and C.~{Caramanis}, ``{Reinforcement Learning for Link Adaptation in
  MIMO-OFDM Wireless Systems},'' in \emph{Proc. IEEE Global Telecommunications
  Conference}, Dec. 2010.

\bibitem{5700414}
M.~{Bennis} and D.~{Niyato}, ``{A Q-learning Based Approach to Interference
  Avoidance in Self-Organized Femtocell Networks},'' in \emph{Proc. IEEE
  Globecom Workshops}, Dec. 2010.

\bibitem{6782491}
J.~{Choi}, ``{Massive MIMO With Joint Power Control},'' \emph{IEEE Wireless
  Communications Letters}, vol.~3, no.~4, pp. 329--332, Aug. 2014.

\bibitem{8415781}
L.~{Zhu}, J.~{Zhang}, Z.~{Xiao}, X.~{Cao}, D.~O. {Wu}, and X.~{Xia}, ``{Joint
  Power Control and Beamforming for Uplink Non-Orthogonal Multiple Access in 5G
  Millimeter-Wave Communications},'' \emph{IEEE Transactions on Wireless
  Communications}, vol.~17, no.~9, pp. 6177--6189, Sep. 2018.

\bibitem{8422442}
C.~{Luo}, J.~{Ji}, Q.~{Wang}, L.~{Yu}, and P.~{Li}, ``{Online Power Control for
  5G Wireless Communications: A Deep Q-Network Approach},'' in \emph{Proc. IEEE
  International Conference on Communications}, May 2018.

\bibitem{725309}
F.~{Rashid-Farrokhi}, L.~{Tassiulas}, and K.~J.~R. {Liu}, ``Joint optimal power
  control and beamforming in wireless networks using antenna arrays,''
  \emph{IEEE Trans. on Commun.}, vol.~46, no.~10, pp. 1313--1324, Oct. 1998.

\bibitem{3gpp36300}
\BIBentryALTinterwordspacing
3GPP, ``{Evolved Universal Terrestrial Radio Access (E-UTRA); Overall
  description},'' {3rd Generation Partnership Project (3GPP)}, TS {36.300},
  Jan. 2019. [Online]. Available:
  \url{http://www.3gpp.org/dynareport/36300.htm}
\BIBentrySTDinterwordspacing

\bibitem{8654723}
R.~{Kim}, Y.~{Kim}, N.~Y. {Yu}, S.~{Kim}, and H.~{Lim}, ``{Online
  Learning-based Downlink Transmission Coordination in Ultra-Dense Millimeter
  Wave Heterogeneous Networks},'' \emph{IEEE Transactions on Wireless
  Communications}, vol.~18, no.~4, pp. 2200--2214, Mar. 2019.

\bibitem{8303773}
S.~Wang, H.~Liu, P.~H. Gomes, and B.~Krishnamachari, ``{Deep Reinforcement
  Learning for Dynamic Multichannel Access in Wireless Networks},'' \emph{IEEE
  Transactions on Cognitive Communications and Networking}, vol.~4, no.~2, pp.
  257--265, Jun. 2018.

\bibitem{8645696}
Y.~{Wang}, M.~{Liu}, J.~{Yang}, and G.~{Gui}, ``{Data-Driven Deep Learning for
  Automatic Modulation Recognition in Cognitive Radios},'' \emph{IEEE
  Transactions on Vehicular Technology}, vol.~68, no.~4, pp. 4074--4077, Apr.
  2019.

\bibitem{8807130}
H.~S. {Jang}, H.~{Lee}, and T.~Q.~S. {Quek}, ``Deep learning-based power
  control for non-orthogonal random access,'' \emph{IEEE Communications
  Letters}, pp. 1--1, Aug. 2019.

\bibitem{8683468}
M.~K. {Sharma}, A.~{Zappone}, M.~{Debbah}, and M.~{Assaad}, ``{Deep Learning
  Based Online Power Control for Large Energy Harvesting Networks},'' in
  \emph{Proceedings of IEEE International Conference on Acoustics, Speech and
  Signal Processing}, May 2019, pp. 8429--8433.

\bibitem{8335785}
W.~{Lee}, M.~{Kim}, and D.~{Cho}, ``Deep power control: Transmit power control
  scheme based on convolutional neural network,'' \emph{IEEE Communications
  Letters}, vol.~22, no.~6, pp. 1276--1279, Jun. 2018.

\bibitem{Alkhateeb2018}
A.~Alkhateeb, S.~Alex, P.~Varkey, Y.~Li, Q.~Qu, and D.~Tujkovic, ``Deep
  learning coordinated beamforming for highly-mobile millimeter wave systems,''
  \emph{IEEE Access}, vol.~6, pp. 37\,328--37\,348, Jun. 2018.

\bibitem{Alrabeiah19}
\BIBentryALTinterwordspacing
M.~{Alrabeiah} and A.~{Alkhateeb}, ``{Deep Learning for TDD and FDD Massive
  MIMO: Mapping Channels in Space and Frequency},'' \emph{Proc. Asilomar
  Conference on Signals, Systems and Computers}, May 2019. [Online]. Available:
  \url{arXiv:1905.03761}
\BIBentrySTDinterwordspacing

\bibitem{8525802}
T.~{Maksymyuk}, J.~{Gazda}, O.~{Yaremko}, and D.~{Nevinskiy}, ``{Deep Learning
  Based Massive MIMO Beamforming for 5G Mobile Network},'' in \emph{Proc. IEEE
  International Symposium on Wireless Systems}, Sep. 2018, pp. 241--244.

\bibitem{framework}
F.~B. Mismar, J.~Choi, and B.~L. Evans, ``{A Framework for Automated Cellular
  Network Tuning with Reinforcement Learning},'' \emph{IEEE Transactions on
  Communications}, vol.~67, no.~10, pp. 7152--7167, oct 2019.

\bibitem{8542687}
P.~{Zhou}, X.~{Fang}, X.~{Wang}, Y.~{Long}, R.~{He}, and X.~{Han}, ``{Deep
  Learning-Based Beam Management and Interference Coordination in Dense mmWave
  Networks},'' \emph{IEEE Transactions on Vehicular Technology}, vol.~68,
  no.~1, pp. 592--603, Jan. 2019.

\bibitem{Xia19}
\BIBentryALTinterwordspacing
W.~Xia, G.~Zheng, Y.~Zhu, J.~Zhang, J.~Wang, and A.~P. Petropulu, ``{A Deep
  Learning Framework for Optimization of MISO Downlink Beamforming},'' Jan.
  2019. [Online]. Available: \url{arXiv:1901.00354}
\BIBentrySTDinterwordspacing

\bibitem{3gpp36213}
\BIBentryALTinterwordspacing
3GPP, ``{Evolved Universal Terrestrial Radio Access (E-UTRA); Physical layer
  procedures},'' {3rd Generation Partnership Project (3GPP)}, TS {36.213}, Dec.
  2015. [Online]. Available: \url{http://www.3gpp.org/dynareport/36213.htm}
\BIBentrySTDinterwordspacing

\bibitem{8665922}
F.~B. {Mismar} and B.~L. {Evans}, ``{Deep Learning in Downlink Coordinated
  Multipoint in New Radio Heterogeneous Networks},'' \emph{IEEE Wireless
  Communications Letters}, vol.~8, no.~4, pp. 1040--1043, Aug. 2019.

\bibitem{Alkhateeb2014}
A.~Alkhateeb, O.~El~Ayach, G.~Leus, and {R. W. Heath Jr.}, ``Channel estimation
  and hybrid precoding for millimeter wave cellular systems,'' \emph{IEEE
  Journal of Selected Topics in Signal Processing}, vol.~8, no.~5, pp.
  831--846, Oct. 2014.

\bibitem{HeathJr2016}
{R. W. Heath Jr.}, N.~Gonzalez-Prelcic, S.~Rangan, W.~Roh, and A.~Sayeed, ``An
  overview of signal processing techniques for millimeter wave {MIMO}
  systems,'' \emph{IEEE Journal of Selected Topics in Signal Processing},
  vol.~10, no.~3, pp. 436--453, April 2016.

\bibitem{Schniter2014}
P.~Schniter and A.~Sayeed, ``{Channel Estimation and Precoder Design for
  Millimeter Wave Communications: The Sparse Way},'' in \emph{Proc. Asilomar
  Conference on Signals, Systems and Computers}, Nov. 2014.

\bibitem{Rappaport2013}
T.~Rappaport, F.~Gutierrez, E.~Ben-Dor, J.~Murdock, Y.~Qiao, and J.~Tamir,
  ``Broadband millimeter-wave propagation measurements and models using
  adaptive-beam antennas for outdoor urban cellular communications,''
  \emph{IEEE Transactions on Antennas and Propagation}, vol.~61, no.~4, pp.
  1850--1859, Apr. 2013.

\bibitem{Rappaport2014}
T.~S. Rappaport, R.~W. Heath~Jr, R.~C. Daniels, and J.~N. Murdock,
  \emph{Millimeter Wave Wireless Communications}.\hskip 1em plus 0.5em minus
  0.4em\relax Pearson Education, 2014.

\bibitem{mnih2013playing}
\BIBentryALTinterwordspacing
V.~Mnih, K.~Kavukcuoglu, D.~Silver, A.~Graves, I.~Antonoglou, D.~Wierstra, and
  M.~Riedmiller, ``{Playing Atari with Deep Reinforcement Learning},''
  \emph{{NIPS Deep Learning Workshop}}, 2013. [Online]. Available:
  \url{http://arxiv.org/abs/1312.5602}
\BIBentrySTDinterwordspacing

\bibitem{Sutton}
R.~S. Sutton and A.~G. Barto, \emph{{Introduction to Reinforcement
  Learning}}.\hskip 1em plus 0.5em minus 0.4em\relax {The MIT Press}, 1998.

\bibitem{goodfellow}
I.~Goodfellow, Y.~Bengio, and A.~Courville, \emph{{Deep Learning}},
  1st~ed.\hskip 1em plus 0.5em minus 0.4em\relax Cambridge, MA, USA: The MIT
  Press, 2016.

\bibitem{1189626}
G.-B. Huang, ``Learning capability and storage capacity of two-hidden-layer
  feedforward networks,'' \emph{IEEE Transactions on Neural Networks}, vol.~14,
  no.~2, pp. 274--281, Mar. 2003.

\bibitem{lin}
L.-J. Lin, ``{Reinforcement Learning for Robots Using Neural Networks},'' Ph.D.
  dissertation, Carnegie-Mellon University, Pittsburg, PA, 1993.

\bibitem{5983301}
M.~{Simsek}, A.~{Czylwik}, A.~{Galindo-Serrano}, and L.~{Giupponi}, ``{Improved
  Decentralized Q-learning Algorithm for Interference Reduction in
  LTE-femtocells},'' in \emph{Proc. Wireless Advanced}, Jun. 2011.

\bibitem{silver}
\BIBentryALTinterwordspacing
D.~Silver, ``{Advanced Topics -- Reinforcement Learning},'' 2015. [Online].
  Available: \url{http://www0.cs.ucl.ac.uk/staff/d.silver/web/Teaching.html}
\BIBentrySTDinterwordspacing

\bibitem{8403587}
F.~B. {Mismar} and B.~L. {Evans}, ``{Partially Blind Handovers for mmWave New
  Radio Aided by Sub-6 GHz LTE Signaling},'' in \emph{{Proc. IEEE International
  Conference on Communications Workshops}}, May 2018.

\bibitem{3gpp38211}
3GPP, ``{NR; Physical channels and modulation},'' {3rd Generation Partnership
  Project (3GPP)}, TS {38.211}, Jun. 2018.

\bibitem{7522613}
A.~I. {Sulyman}, A.~{Alwarafy}, G.~R. {MacCartney}, T.~S. {Rappaport}, and
  A.~{Alsanie}, ``{Directional Radio Propagation Path Loss Models for
  Millimeter-Wave Wireless Networks in the 28-, 60-, and 73-GHz Bands},''
  \emph{IEEE Transactions on Wireless Communications}, vol.~15, no.~10, pp.
  6939--6947, Oct. 2016.

\bibitem{6932503}
T.~{Bai} and R.~W. {Heath Jr.}, ``{Coverage and Rate Analysis for
  Millimeter-Wave Cellular Networks},'' \emph{{IEEE Transactions on Wireless
  Communications}}, vol.~14, no.~2, pp. 1100--1114, Feb. 2015.

\bibitem{my_jbpcic_code}
\BIBentryALTinterwordspacing
F.~B. Mismar. (2019) Source code. [Online]. Available:
  \url{{https://github.com/farismismar/Deep-Reinforcement-Learning-for-5G-Networks}}
\BIBentrySTDinterwordspacing

\end{thebibliography}

\end{document}